\documentclass{aa} 
\usepackage{graphicx}
\usepackage{txfonts}
\usepackage[colorlinks=true,linkcolor=blue,citecolor=blue, filecolor=blue, urlcolor=blue]{hyperref}
\begin{document}

\title{An interferometric search for SiO maser emission in planetary nebulae}

   \author{Rold\'an A. Cala \inst{1}
          \and          
          Jos\'e F. G\'omez \inst{1}
          \and
          Luis F. Miranda \inst{1}
          \and
          Hiroshi Imai \inst{2,3}
          \and 
          Itziar de Gregorio-Monsalvo \inst{4}
           \and
           \\
          Florin Placinta Mitrea\inst{1}
         \and
          Mayra Osorio \inst{1}
          \and
          Guillem Anglada \inst{1}
          }

   \institute{Instituto de Astrof\'{\i}sica de Andaluc\'{\i}a, CSIC, 	Glorieta de la Astronom\'{\i}a s/n, E-18008 Granada, Spain\\
              \email{rcala@iaa.es}
         \and
         Center for General Education, Comprehensive Institute of Education, Kagoshima University, 1-21-30 Korimoto, Kagoshima 890-0065, Japan
         \and
         Amanogawa Galaxy Astronomy Research Center,
Graduate School of Science and Engineering, Kagoshima University, 1-21-30 Korimoto, Kagoshima 890- 0065, Japan
\and 
European Southern Observatory, Alonso de Cordova 3107, Casilla 19, Vitacura, Santiago, Chile
             }

   \date{Received ; accepted }

 
  \abstract
   {Maser emission of SiO, H$_2$O and, OH is widespread in Asymptotic Giant Branch (AGB) stars with oxygen(O)-rich envelopes. This emission quickly disappear during the post-AGB phase and is extremely rare in planetary nebulae (PN). So far, only eight PNe have been confirmed to show OH and/or H$_2$O maser emission, and none has ever been found to show SiO maser emission.}
   {We intend to obtain the first detection of a SiO maser from a PNe.  Such a detection would provide us with a useful tool to probe mass-loss in PNe at a scales of a few AU from the central star, much shorter than the scales traced by H$_2$ or OH masers.}
   {We compiled two different samples. The first one comprises all known PNe with confirmed OH and/or H$_2$O maser emission, as well as two candidate PNe showing OH masers. For the second sample we compiled single-dish SiO maser detections in the literature, and compared them with catalogs of PNe and radio continuum emission (which could trace photoionized gas in a PNe). We identified five targets (either PN or radio continuum sources) within the beam of the single-dish SiO maser observations. We then carried out interferometric observations of both samples with the Australia Telescope Compact Array, to confirm the spatial association between continuum and SiO maser emission.}
   {{ We found no SiO maser emission associated with any}  confirmed or candidate PN. In all targets, except IRAS 17390$-$3014, there is no spatial coincidence between SiO masers and radio continuum emission. While in IRAS 17390$-$3014 we cannot completely rule out a possible association, it is unlikely that the radio continuum emission arises from a planetary nebula.}
   {The absence of SiO maser emission in PNe showing OH or H$_2$O masers is of special interest, since thermal SiO emission has been reported in at least one of these targets, indicating that SiO molecules can be present in gas phase. Since some maser-emitting PNe show evidence of having O-rich outer envelopes, and carbon(C)-rich central stars and inner envelopes, we speculate that SiO abundance could be very low in the central regions where physical conditions are optimal for maser pumping, and C-bearing molecules may be dominant in gas phase at those locations.}

   \keywords{Masers -- stars: AGB and post-AGB -- stars: winds, outflows -- planetary nebulae: general -- radio continuum: stars
               }

\titlerunning{SiO maser emission in planetary nebulae}
\authorrunning{R.~A. Cala et al.}
   \maketitle 
%

\section{Introduction}

Circumstellar maser emission of the SiO, H$_2$O, and OH molecules is widespread in Asymptotic Giant Branch (AGB) stars with O-rich envelopes. These different species show a stratified distribution, with SiO masers located close to the star $\la 10$ au \citep{dia94,des00,rei07}, and H$_2$O, and OH farther away { \citep[$\simeq 10-10^2$ and $10^3 -10^4$ au, respectively; e.g.,][]{bow94,ric12,rei77,eto10}}, tracing a nearly spherical expansion of the envelope at different radii. This maser emission tends to disappear after the end of the AGB phase, when mass-loss drops by several orders of
magnitude, from rates up to $\simeq 10^{-4}$ M$_\odot$ yr$^{-1}$ in the AGB \citep{vas93,blo95a} to $10^{-7}-10^{-8}$ M$_\odot$ yr$^{-1}$ 
in the post-AGB phase \citep{vas94,blo95b}. 

Considering the estimated timescale of disappearance of these maser species after the end of the AGB \citep[10, 100, 1000 yr for SiO, H$_2$O, and OH, respectively;][]{gom90,eng07}, and the duration of the transitional post-AGB phase \citep[$10^3-10^5$ yr,][]{mil16} no SiO nor H$_2$O maser emission was expected when the planetary nebula  (PN) phase begins { (i.e., when the central star becomes hot enough, $T_{\rm eff}\simeq 25000$~K, to photoionize its circumstellar environment)}, while only extremely young PNe { (shortly after the onset of this photoionization)} should show OH masers. 
This makes OH-maser emitting PNe \citep[OHPNe,][]{zij89} very interesting objects to understand the early evolution of PNe. There are only six confirmed OHPNe \citep{usc12,qia16b}, and several additional candidates \citep{cal22}.

Despite not being expected, five PNe are known to emit H$_2$O masers \citep{mir01,degr04,gom08,usc14,gom15}. The first identified case, K 3-35 \citep{mir01}, actually emits both OH and H$_2$O masers. Two other PNe,   IRAS 17347-3139 \citep{degr04,taf09} and IRAS 16333-4807 \citep{usc14,qia16b}, also show emission of both species. The unexpected detection of H$_2$O masers suggests that this emission in
PNe is not the remnant of that seen in the AGB phase, which is related to isotropic winds, but traces non-spherical mass-loss events instead \citep{sua07}. In particular, interferometric observations usually show that OH and H$_2$O maser emission in PNe is associated with 
 equatorial ejections (expanding tori) and/or collimated jets \citep{mir01,usc14,gom15,qia16b}. While these mass-loss phenomena allow the presence of maser emission in PNe for longer than previously expected, { it is generally assumed} that only extremely young PN can be maser emitters, since the advance of the ionization front would eventually photodissociate OH and H$_2$O molecules at the locations where those masers can be produced. The assumption of extreme youth of maser-emitting PNe is compatible with their scarcity.
 
Considering that new, non-spherical mass-loss events in the post-AGB and PN phases can provide the conditions of population inversion of OH and H$_2$O levels (required for maser emission), it is reasonable to think that the same might happen in the case of SiO.
In fact, a few late AGB and post-AGB stars are known to harbor SiO maser emission \citep[e.g.,][]{yoo14,nak03a,ama22}, which is  probably tracing either the base of collimated outflows \citep{ima05} or circumstellar toroids \citep{san02}. However, in the case of objects already in the PN phase, several attempts have been made to detect SiO masers, all with negative results \citep[e.g.,][]{bac97,nym98}.  In any case, SiO searches specifically targeting PNe remain scarce.

Obtaining a first detection of SiO masers in PNe would open a new window of research. It would allow us to probe mass-loss at scales different from those traced by H$_2$O and OH, probably closer to the central star. Moreover, we could characterize the physical conditions that favor the presence of SiO masers, and how it correlates with the presence of other maser species. 

In this paper we present the results of a new sensitive survey for SiO masers in confirmed and candidate PNe, carried out with the Australia Telescope Compact Array (ATCA). We selected two different samples, which are described in Sec. \ref{sec:samples}. We present our results in Sec. \ref{sec:results}, which we subsequently discuss in Sec. \ref{sec:discussion}

\section{Source samples}
\label{sec:samples}

\subsection{Sample 1: planetary nebulae with known OH and/or H$_2$O masers}
\label{sec:sample1}

PNe with either OH or H$_2$O maser emission are considered to be at the beginning of this evolutionary phase. Thus, these nascent PNe are among the best targets to search for SiO maser emission as well, since we expect it to disappear shortly after the end of the AGB phase, at even shorter timescales than the OH or H$_2$O masers. Our sample comprises 11 sources and includes all eight confirmed PN with OH or H$_2$O maser emission, as well as two additional objects whose nature as PN has not been confirmed, presented in \cite{cal22}: IRAS 16372$-$4808 and OH 341.6811+00.2634. We also included NGC 6302, a well known PN displaying OH line emission \citep{tur79,zij89}, although it has been suggested that this emission could be thermal, instead of maser \citep{gom16,qia20}. The list of sources is given in the first part of Table \ref{tab:obs_atca}.

Some of these objects have already been observed in SiO transitions at 43 GHz \citep{jew91,nym98}, but with modest sensitivities (rms noise $\simeq 1$ Jy). The observations in this paper are significantly more sensitive (around two orders of magnitude).

   \begin{table*}
      \caption[]{ Parameters of our ATCA observations.}
      \label{tab:obs_atca}
      \centering                          
\begin{tabular}{llllllll}     
\hline\hline                 
Source & R.A.(J2000) & Dec(J2000) & gain calibrator & beam size & beam p.a. & $T_{\rm on}$ & References\\    
	   &			& 			 & 				   & (arcsec)  & (degree) & (hour) \\
\hline 
\multicolumn{8}{c}{Sample 1} \\
\hline     
IRAS 15103$-$5754 & 15:14:18.397 & $-58$:05:20.47 &  PMN J1515-5559 & $7.2\times 4.9$ & $+85$ & 0.65 & 1,2\\
IRAS 16333$-$4807 &  16:37:06.601 & $-48$:13:42.88  &  IERS B1646-506 & 
$8.0\times 4.6$ & $+78$ & 0.65 & 3,4\\ 
IRAS 16372$-$4808 & 16:40:55.818 & $-48$:13:59.93 & IERS B1646-506 & $7.1\times 4.6$ & $+82$ & 0.65 & 5,6 \\ 
OH 341.6811+00.2634 & 16:51:53.640 & $-43$:47:16.50 & IERS B1646-506 & $9.5\times 4.1$ & $+81$ & 0.65 & 5 \\ 
NGC 6302 & 17:13:44.489 & $-37$:06:11.71 & IERS B1729-373 & $6.9\times 4.3$  & $+75$ & 0.65 & 7,8\\  
IRAS 17347$-$3139 & 17:38:00.624  & $-31$:40:54.91 & IERS B1741-312 & 
$6.8\times 4.2$	& $+75$ & 0.66 & 9,10 \\ 
JaSt 23 & 17:40:23.067 & $-27$:49:12.00 & IERS B1741-312 &
$6.5\times 4.1$ & $+88$ & 0.67 & 11,12 \\ 
IRAS 17393$-$2727 & 17:42:33.140 & $-27$:28:24.70 & IERS B1741-312 & 
$7.2\times 3.9$ & $-88$ & 0.65 & 13,14\\ 
IRAS 18061$-$2505 & 18 09 12.400 & $-25$:04:34.50 & IERS B1817-254 & $6.2\times 4.1$ & $+72$ & 0.65 & 15,16 \\ 
Vy 2-2 & 19:24:22.218 & $+09$:53:56.33 & IVS B1936+046	& 
$6.8\times 5.7$ & $+78$ & 1.24 & 17,18\\ 
K 3-35 & 19:27:44.026 & $+21$:30:03.57 & IVS B1936+046	&
$8.1\times 5.1$ & $+31$ & 0.66 & 19,20\\
\hline
\multicolumn{8}{c}{Sample 2} \\
\hline
IRAS 17239$-$2812 &  17:27:05.030  & $-28$:15:27.50 & PKS 1710-269 & $0.76\times 0.18$ & $+7$ & 0.74 & 21,22\\
IRAS 17390$-$3014 & 17:42:18.870 & $-30$:15:29.00 & PKS 1710-269 & 
$0.73\times 0.17$ &	$+6$	& 0.74 & 23,24\\ 
H 2-18 & 17:43:28.840 & $-21$:09:52.50 & PKS 1710-269 & $0.95\times 0.18$ & $+3$ & 0.74 & 25,26\\
IRAS 18052$-$2016 & 18:08:16.380 &  $-20$:16:11.30 & IERS B1829-207 &  $1.37\times 0.17$ & $+4$& 0.74 & 27,28\\
IRAS 19508$+$2659 & 19:52:57.900 &  $+27$:07:44.70 & VCS2 J1956+2820 & $1.12\times 0.16$ & $+2$ & 0.57 & 29,30\\
\hline                                   
\end{tabular}
\tablefoot{Right ascension and declination represent the coordinates of the phase center of the observations. ``Gain calibrator'' is the complex-gain calibrator used for each target. Beam sizes (full width at half maximum) and beam position angles (p.a., measured north to east) are those of the continuum images at 43 GHz. $T_{\rm on}$ is the total integration time on each source.
The coordinates of H 2-18 are those given in \citet{gat83}, since the position given in the SIMBAD database, at the time of { our observations, actually corresponded} to a different object, IRAS 17403-2107 (see section \ref{h2-18}). { Sample 1 comprises PNe with known OH and/or H$_2$O masers (Section \ref{sec:sample1}). The references listed for this sample correspond to their identification as PNe and the report on the presence maser emission. Sample 2 includes single-dish SiO maser detections near PNe and radio continuum sources (Section \ref{sec:sample2}). Their references correspond to their identification as PNe or the presence of radio continuum emission, and the single-dish reporting of maser emission.}
\\ References: 
(1) \cite{gom15}, (2) \cite{sua15}, (3) \cite{kim01}, (4) \cite{usc14}, (5) \cite{cal22}, (6) \cite{sev97b}, (7) \cite{pay88},  (8) \cite{hen67}, (9) \cite{degr04}, (10) \cite{gom05}, (11) \cite{jast04}, (12) \cite{sev97a},  (13) \cite{pot87}, (14) \cite{gar07}, (15) \cite{sua06}, (16) \cite{gom08}, (17) \cite{vys45}, (18) \cite{sea83}, (19) \cite{mir98}, (20) \cite{mir01}, (21) \cite{hal90}, (22) \cite{dou96}, (23) \cite{deg04}, (24) \cite{con98}, (25) \cite{gat83}, (26) \cite{izu95}, (27) \cite{dea07}, (28) \cite{gar88}, (29) \cite{whi05}, (30) \cite{nak03a}
}
  \end{table*}

\subsection{Sample 2: previous single-dish SiO maser detections near planetary nebulae and radio continuum sources}
\label{sec:sample2}

We carried out an extensive search in the literature and in publicly available catalogs for, on the one hand, reports on single-dish detections of SiO maser emission and, on the other hand, different catalogs of PNe and radio continuum emission, considering the latter as a possible tracer of free-free radiation from ionized gas in a PN{ (the list of references for the published works and catalogs we compiled would be too long to be detailed here, but some examples can be seen in references 21 to 30 in Table \ref{tab:obs_atca}, and in the description of each source below)}. We then cross-matched these single-dish SiO detections with PNe and radio continuum sources, which resulted in five candidate sources (Table \ref{tab:obs_atca}, bottom five sources) { within the beam of the single-dish SiO observations}. A definite confirmation of the association of maser emission with known PNe or with radio continuum emission requires interferometric confirmation, to ascertain that the SiO maser emission actually arises from the target object, and not from a different source within the single-dish beam.

\section{Observations and data processing}

\label{sec:obs}

ATCA observations were carried out in two sessions. 
The sample described in section \ref{sec:sample1} (Sample 1, project C3324, PI: R. Cala) was observed on 2019 October 21, with the H168 configuration of ATCA (minimum and maximum baselines of 61 and 192 m, respectively, when discarding antenna CA06)
and the sample in section \ref{sec:sample2} (Sample 2, project C3391, PI: R. Cala) on 2020 October 16 with the 6B configuration (baselines between 214 and 5969 m). 
In both sessions, we observed at two different intermediate frequencies (IFs), centered at 43 and 45 GHz, obtaining { full linear polarization products} with a  bandwidth of 2 GHz each. We used the CFB 64M-32k mode of the Compact Array Broadband Backend (CABB) of ATCA, which samples { each 2-GHz bandwidth into broadband channels with a width of 64 MHz}, used for continuum observations. { The CABB internally oversamples the total bandwidth, by obtaining broadband channels with separation of half the channel width, but only half of them (with final channel separation equal to the channel width) are provided to the user for the continuum dataset (a total of 32 channels of 64 MHz for the CFB 64M-32k mode).} Five of the { internal broadband} channels were independently zoomed in { with a finer spectral resolution, to obtain five individual spectral windows}, each including one SiO transition and sampled into 2048 channels of 32 kHz width (velocity resolution 0.22 km s$^{-1}$). Of these channels, only the central 1847  ones were kept, providing a total coverage of $\simeq 404$ km s$^{-1}$.  An additional set of 5 adjacent { internal broadband} channels was combined and also zoomed in, providing a spectral window with 6144 channels, of which the central 5943 ones were kept (velocity resolution and coverage of 0.22 and  1296 km s$^{-1}$, respectively). This last set of channels  includes both the hydrogen recombination line H53$\alpha$ (rest frequency = 42951.968 MHz), and the  $^{29}$Si$^{16}$O ($v$=0, J=1-0) line. 
Therefore, in total we observed six SiO transitions: four of the main { isotopologue, $^{28}$Si$^{16}$O (hereafter $^{28}$SiO), and two of $^{29}$Si$^{16}$O} (hereafter $^{29}$SiO). These transitions are listed in Table \ref{tab:restfreq}. 

In both sessions, the bandpass calibrator was 3C 279 (PKS 1253-055), while the absolute flux calibrator was PKS 1934-638.
Further information on our observational parameters, including the complex-gain calibrator for each target, is listed in Table \ref{tab:obs_atca}. We  note that the observations of project C3324 (sample 1) were carried out in the H168 configuration, in which five of the ATCA antennas have a relatively compact distribution, with maximum baseline of 192 m, and there is a sixth antenna (CA06) at a fixed position, $\simeq 4.5$ km away from the main array. Although we included CA06 in our initial maps, the quality of the resulting maps  was poor, given the highly inhomogeneous sampling of the uv-plane. Therefore, we excluded antenna CA06 from the data we presented for this project C3324. However, for project C3391 (sample 2) we used the whole array of six antennas, which gave good-quality maps, since the antenna distribution was optimized for observations at high angular resolution.

   \begin{table}
      \caption[]{Observed SiO transitions.}
      \label{tab:restfreq}
      \centering                          
\begin{tabular}{c c c}     
\hline\hline                 
{ Isotopologue} & Transition & rest frequency \\    
		   &			& (MHz) \\
\hline                        
$^{28}$Si$^{16}$O&  $v$=0, J=1-0 & 43423.853 \\      
$^{28}$Si$^{16}$O&  $v$=1, J=1-0 & 43122.090 \\
$^{28}$Si$^{16}$O&  $v$=2, J=1-0 & 42820.570 \\
$^{28}$Si$^{16}$O&  $v$=3, J=1-0 & 42519.375\\
$^{29}$Si$^{16}$O&  $v$=0, J=1-0 & 42879.941\\
$^{29}$Si$^{16}$O&  $v$=1, J=1-0 & 42583.830\\
\hline                                   
\end{tabular}
\tablefoot{These rest frequencies were taken from the \citet{lov04} list. In the case of the $^{29}$Si$^{16}$O, $v$=1, J=1-0, for which no frequency is available in that list, we assumed the value given by \citet{mul05}.}
  \end{table}

Data were calibrated with the Miriad software \citep{sau95}, following standard procedures for millimeter data processing in the ATCA users' guide. We then further processed the calibrated data using both the Common Astronomy Software Applications \citep[CASA, ][]{cas22} and the Astronomical Image Processing System \citep[AIPS,][]{fom81}. We used CASA for initial data inspection, creating images and data cubes with task tclean. With this, we could identify the sources where line and continuum emission were detected. We also attempted self-calibration of the data with CASA, but its performance proved to be somewhat worse than AIPS self-calibration, at least for our data, using equivalent parameters. As an example, the signal-to-noise improvement of AIPS vs CASA was a factor of 2  when self-calibration was carried out with continuum data (sample 1), and a factor of $\simeq 1.5$ when using masers (sample 2). { Even though this improvement may seem modest, it could be important for a detection project like ours, in case the emission is close the sensitivity limit. Therefore}, we decided to continue all further data processing using AIPS. Tasks mentioned below correspond to the AIPS package, unless stated otherwise.

Since the ATCA { does not correct the frequency drift due to the Earth's motion during the observing session,
we carried out an off-line correction of the visibilities of the line spectral windows for this Doppler shift}, using task CVEL, to align all data to the same velocity (with respect to the kinematical definition of the Local Standard of Rest, hereafter LSR). { All radial velocities mentioned in this paper are given with respect to the LSR.} In the fields where obvious SiO emission was detected (most sources in sample 2), we selected the velocity channel with strongest emission of the $^{28}$SiO $v$=1 line, and carried out phase and amplitude self-calibration. In most cases we could obtain solutions at intervals as short as the integration cycle time of the individual visibilities (6 seconds for sample 1,  10 seconds for sample 2). We then copied the solutions to the whole dataset, including all spectral lines and the continuum emission. For sample 1, where no SiO emission was detected, we carried out self-calibration on the radio continuum data at 43 GHz,  and the solutions were then copied to the spectral line data and the continuum at 45 GHz. { By using the same self-calibration solutions for all data in each source, we ensure a better relative accuracy in the flux density measurements.}

{ Moreover, for sample 1, the sampling of the interferometric plane provided by the ATCA with the short time span covered by the observations of each source ($\simeq 45$ min in most cases) is poor, which results in point spread functions with strong sidelobes. In the case of the unresolved sources with a good signal-to-noise ratio, this is not a major problem, since the CLEAN algorithm can work properly and remove the sidelobes. However, the continuum emission in NGC 6302 is resolved, which significantly affects the image reconstruction. In fact, several of the resulting sidelobes of the maps are actually stronger than the peak closest to the expected location of the source. Even restricting deconvolution to this central peak, we could not obtain continuum maps with good enough fidelity to provide reliable source parameters. The interferometric visibilities, however, clearly show the presence of emission. Thus, the parameters of the continuum emission in this source, presented below, were obtained by fitting an elliptical Gaussian function to the visibilities using task uvmodelfit of CASA.}

We subtracted the continuum emission from the visibilities of the spectral lines using task UVLIN, by selecting channels without line emission in the spectral window of each transition. For the continuum emission, we used the broadband channels covering 2 GHz at each IF, flagging out the frequencies were the line emission was detected.
Final maps were obtained with task IMAGR, using Brigg's weighting of visibilities, with robust parameter equal 0 (which is equivalent to 0.5 in the imaging tasks of CASA, such as tclean). The continuum images were obtained with multi-frequency synthesis. All continuum and line images were deconvolved using the CLEAN algorithm \citep{cla80}, implemented in task IMAGR.

The absolute astrometric accuracy of the ATCA has been estimated to be $\simeq 0.4''$  for well-calibrated data and under good weather conditions \citep[e.g.,][]{cas97,ell18}. Since we self-calibrated our data and the weather conditions were acceptable but not excellent, our astrometric accuracy is then determined by the positional accuracy of our model for each source (the position of the radio continuum emission at 43 GHz for sample 1 or the peak of the $^{28}$SiO ($v$=1, J=1-0) emission in sample 2). A comparison of the position in final images of the features used for self-calibration,  with that of their counterpart sources in the Gaia DR3 catalog \citep{gaia16,gaia23}, 2MASS point source catalog \citep{2mass}, and Spitzer catalogs \citep{spitzer} allows us to estimate that our absolute astrometry is accurate to $<1.2''$ for sample 1, and $<0.7''$ for sample 2. 
Positional information in our data was obtained by fitting elliptical Gaussians to the emission, with task JMFIT. The positions we report in the results of this paper correspond to the emission used as a self-calibration reference. Therefore, for sources detected in continuum emission, we provide the absolute positions at 43 GHz. In the case of sources with detected maser emission, considering the angular resolution of the observations and the signal-to-noise ratio achieved,  the different maser components of all SiO transitions within each source were compatible, within the errors, so we cannot provide information about the spatial distribution of the emission. Therefore, for these maser detections, we only give the estimated position of the strongest maser component for the $^{28}$SiO($v=1$, J=1-0) transition as representative of all SiO maser emission, in the results presented in Section \ref{sec:results2}.

The uncertainty of flux calibration in ATCA observations at 7 mm is estimated to be $\simeq 10$\%. However, we note that we do not include this uncertainty when quoting flux density errors in the tables of this paper. The flux density errors in this paper only consider the contribution of thermal noise in the maps, and are given at a 1$\sigma$ level. These errors provide a good representation of the relative flux uncertainties among the different SiO transitions, and between these and the continuum emission, since they all share the same flux and complex gain calibration, and observations are simultaneous. The errors we quote are thus useful for estimating line ratios and their uncertainties in our data. However, the 10\% uncertainty in absolute calibration should be added to the flux density errors quoted in this paper, when comparing with the emission from other datasets. We also note that the errors quoted in this paper for all parameters are given at a $1\sigma$ level, while upper limits are at $3\sigma$.

\section{Results}
\label{sec:results}

We have not detected any SiO maser emission associated with PNe or PN candidates, in either observed sample. { In particular, in the young PNe of sample 1 we detected radio continuum emission but no SiO. In sample 2, we detected SiO emission that corresponds { that} previously reported  in single-dish observations toward these objects, but we could not confirm any association of SiO and radio continuum emission. Instead, our interferometric observations allow us to firmly establish the association of the detected SiO emission with strong, point-like infrared sources. Together with the absence of radio continuum emission, we infer that these SiO-emitting sources are most likely to be AGB or post-AGB stars. }

\subsection{Sample 1}

In the case of the sample of confirmed and candidate PNe that emit maser emission of other molecular species (OH and/or H$_2$O), we do not detect any SiO emission within the observed fields. We detect the continuum emission associated with all the targets in this sample. 
{ A summary of the results is presented in Table \ref{tab:results1}. The continuum emission  of all sources, except NGC 6302, is unresolved in our images (with an angular resolution $\simeq 6''$).
 As mentioned in section \ref{sec:obs}, the position and flux densities reported in Table \ref{tab:results1} for NGC 6302 were obtained by fitting a model of an elliptical Gaussian source to the visibilities. The full width at half maximum of the fitted Gaussian was $11''\times 9''$, with position angle $\simeq -53^\circ$.}

\begin{table*}
\caption{Radio continuum emission { and representative SiO upper limits} in confirmed and candidate PNe with OH and/or H$_2$O masers (sample 1)}             
\label{tab:results1}      
\centering          
\begin{tabular}{lllllrr}     
\hline\hline       
Source & R.A.(J2000) & Dec (J2000) & $S_{43}$ &  $S_{45}$ & $S_{v1}$ & $S_{v2}$\\
		&			&				& (mJy)	  &  (mJy)	& (mJy)	  &  (mJy)	  \\
\hline         
IRAS 15103$-$5754 & 15:14:18.42 &  $-58$:05:21.0 
& $18.41\pm 0.09$ & $19.26\pm 0.15$ & $<8$ & $<8$ \\ 
IRAS 16333$-$4807 & 16:37:06.55 &  $-48$:13:43.9
& $103.6\pm 0.3$ & $109.9\pm 0.7$ & $<8$ & $<8$\\
IRAS 16372$-$4808 & 16:40:55.97 &  $-48$:13:59.9
& $57.98\pm 0.08$ & { $59.6\pm 0.4$} & $<7$ & $<6$\\ 
OH 341.6811+00.2634 & 16:51:53.66 & $-43$:47:16.8 & $16.44\pm 0.08$ & { $16.32\pm 0.11$} & $<7$ &$ <7$\\ 
NGC 6302 & { 17:13:44.56} & { $-37$:06:10.4} &
$1475.2\pm 1.1$ & $1449.2\pm 1.2$ & $<41$ & $<43$\\
IRAS 17347$-$3139 & 17:38:00.55 & $-31$:40:55.4 & $320.15\pm 0.15$ & $313.3\pm 2.2$ & $<7$ &$ <5$\\
JaSt 23 & 17:40:23.05 &  $-27$:49:11.5 & $21.98\pm 0.05$ & { $22.49\pm 0.15$} & $<7$ & $<6$\\
IRAS 17393$-$2727 & 17:42:33.13 & $-27$:28:25.0 & $22.41\pm 0.08$ &
{ $23.38\pm 0.14$} & $<6$ & $<6$\\ 
IRAS 18061$-$2505 &  18:09:12.44 &  $-25$:04:34.1 &  $55.88\pm 0.07$ & { $56.7\pm 0.6$} & $<6$ & $<5$\\ 
Vy 2-2 & 19 24 22.22 & $+09$:53:56.3 & $253.28\pm 0.14$ & { $264.0\pm 1.1$} & $<7$ & $<7$\\
K 3-35 & 19:27:44.01 & $+21$:30:05.1  & $12.45\pm 0.07$ & $12.38\pm 0.14$ & $<7$ & $<8$ \\ 
\hline
\end{tabular}
\tablefoot{The positions listed in this table were obtained by fitting elliptical Gaussians to the images at 43 GHz. $S_{43}$ are  $S_{45}$ the flux densities at 43 and 45 GHz, respectively. { $S_{v1}$ and $S_{v2}$ are the 3$\sigma$ upper limits per channel of $v=1$ and 2 of $^{28}$SiO, respectively, as representative of the sensitivity for spectral line detection in our data}. Uncertainties are at $1\sigma$ level. In the case of NGC 6302, for which we could not obtain { reliable images, the reported position and flux densities of the continuum emission were obtained by fitting a model of elliptical Gaussian source to the visibilities. For this source, we report the formal error for the flux density obtained in the fit, which most likely underestimate the actual flux density uncertainty if the flux distribution significantly departs from a Gaussian function. The SiO upper limits for NGC 6302 were calculated in spectra obtained directly from the calibrated visibility data.}}
\end{table*}

We note that values of the radio continuum emission at 43 and 45 GHz using this dataset were already presented for IRAS 16372$-$4808 and OH 341.6811+00.2634 by \cite{cal22}. Here we present slightly revised values for these sources. { The main difference with respect to \cite{cal22} is that, in that paper, the data at each frequency were self-calibrated independently, while in the present paper we used the same self-calibration solutions (those obtained at 43 GHz), as mentioned in Section \ref{sec:obs}.} The new values are compatible with the previously reported ones.

{ We also detected the H53$\alpha$ recombination line in seven of the targets of this sample. These results will be presented in a forthcoming paper. }

\subsection{Sample 2}

\label{sec:results2}

SiO maser emission was detected in most of the observed fields (Table \ref{tab:results2}). { SiO maser emission had been previously reported in those sources} via single-dish observations, but we provide accurate interferometric positions, for the first time in most cases. Based on this, in no case we could confirm that it was associated with a PN. In particular, the position of the SiO emission we obtain in our interferometric maps does not match any of the radio continuum sources reported in their neighborhood or, in some cases, the reported continuum sources may be spurious. Only in the case of IRAS 17390$-$3014, there is an extended radio continuum source that { overlaps the optical/infrared position of the SiO-emitting object}. However, as discussed in \ref{sec:17390}, we think it unlikely that the radio continuum source corresponds to a planetary nebula. { Moreover, the large offsets found in some sources between SiO and the previously reported radio continuum emission cannot be due to proper motions, since these are lower than 7 mas in all cases, according to Gaia data.} We present below the results for each individual source, discussing the relative position of the SiO emission with respect to radio, infrared, and optical sources that could be potential hosts of these masers.

\begin{table*}
\caption{SiO(J=1-0) emission for sources of sample 2}             
\label{tab:results2}      
\centering          
\begin{tabular}{lrlllllll}     
\hline\hline       
Source 				& Transition 	& R.A. (J2000) & Dec (J2000) &  $S_{\rm peak}$ 	&  $V_{\rm peak}$  & $\int S_\nu dV$ & $V_{\rm min}$ & $V_{\rm max}$ \\
					&				&		 	& 				&  (mJy)	  & (km s$^{-1}$) & (mJy km s$^{-1}$) &   (km s$^{-1}$) &  (km s$^{-1}$) \\
\hline         
IRAS 17239$-$2812 	& $^{28}$SiO,$v=0$ 		&  &  	& $146\pm 3$  & $-27.6$ 			& $377\pm 14$ 			& $-30.2$ 		& $-27.2$ \\
					& $^{28}$SiO,$v=1$ 		& 17:27:05.41   & $-28$:15:29.8 & $7578\pm 5$ & 	$-29.6$			& $21436\pm 25$			& $-37.6$		& $-23.0$ \\ 
					& $^{28}$SiO,$v=2$		&   & 	& $9627\pm 8$ & $-29.8$			& $28950\pm 30$ 			& $-38.3$ 		& $-23.5$ \\
					& $^{28}$SiO,$v=3$		&   & 	 & $1642\pm 5$ & $-28.6$			& $2885\pm 18$ 			& $-31.7$ 		& $-26.7$ \\
					& $^{29}$SiO,$v=0$ & &	& $490\pm 4$ 	& $-24.8	$ 		& $1648\pm 21$ &  $-30.3$ & $-23.5$ \\
					& $^{29}$SiO,$v=1$ & &	& $<18$\\
IRAS 17390$-$3014	& $^{28}$SiO,$v=0$ 		&  &  & $85\pm 6$ & $-23.9$			& $76\pm 9$ 			& $-24.8$		& $-23.7$ \\
					& $^{28}$SiO,$v=1$ 		& 17:42:18.86 &  $-30$:15:29.0  	& $2128\pm 5$ & $-20.4$			& $5310\pm 30$			& $-27.1$		& $-18.0$ \\ 
					& $^{28}$SiO,$v=2$		&  & 	& $4538\pm 8$ & $-20.8$			& $9780\pm 30$ 			& $-27.3$ 		& $-17.1$ \\
					& $^{28}$SiO,$v=3$		&  & 	& $739\pm 5$  & $-20.6$			& $845\pm 12$ 			& $-22.2$		& $-20.0$ \\
					& $^{29}$SiO,$v=0$ & &	& $<15$		\\
					& $^{29}$SiO,$v=1$ & &	& $<18$\\
H 2-18				& $^{28}$SiO,$v=0$		&				&					& $<25$\\
					& $^{28}$SiO,$v=1$ 		&				&					& $<45$\\
					& $^{28}$SiO,$v=2$ 		&				&					& $<25$\\
					& $^{28}$SiO,$v=3$ 		&				&					& $<25$\\
					& $^{29}$SiO,$v=0$ & &	& $<48$		\\
					& $^{29}$SiO,$v=1$ & &	& $<15$ \\

IRAS 18052$-$2016	& $^{28}$SiO,$v=0$ 		&  &  	& $60\pm 6$			& 	$+52.3$	& $200\pm 30$ & $+50.5$ & $+63.3$\\
					& $^{28}$SiO,$v=1$ 		&  18:08:16.39 &  $-20$:16:12.2 	& $4424\pm 4$ & $+50.5$			& $30130\pm 70$ 			& $+17.4$		& 	$+59.2$\\ 
					& $^{28}$SiO,$v=2$		&  & 	& $3866\pm 6$ & $+49.9$			& $18180\pm 60$ 			& $+33.3$		&	$+61.5$\\
					& $^{28}$SiO,$v=3$		&  &  	& $133\pm 5$  & $+54.7$			& $350\pm 25$ 			& $+49.2$		&	$+55.6$\\
					& $^{29}$SiO,$v=0$ & &	& $317\pm 5$  & $+52.8$ 			& $928\pm 24$			& $+49.1$		& $+56.1$\\
					& $^{29}$SiO,$v=1$ & &	& $<18$\\
IRAS 19508+2659		& $^{28}$SiO,$v=0$		& 				&					& $<15$\\
					& $^{28}$SiO,$v=1$		& 19:52:57.85 & $+27$:07:45.4	& $927\pm 10$ & $+8.6$			& $3230\pm 30$ 			& $+2.54$			& $+10.1$\\ 
					& $^{28}$SiO,$v=2$		&  & 	& $897\pm 10$ & $+3.3$			& $3550\pm 30$ 			& $+2.0$			& $+9.9$\\
					& $^{28}$SiO,$v=3$		&				&					& $<20$ \\
						& $^{29}$SiO,$v=0$ & &	& $<16$\\
						& $^{29}$SiO,$v=1$ & &	& $<25$\\	
\hline
\end{tabular}
\tablefoot{ For each source, we give the position of the peak emission of the  $^{28}$SiO ($v$=1, J=1-0) transition, which we used as a reference for self-calibration. The positions of all other velocity components and transitions are compatible with this one, within the errors. 
$S_{\rm peak}$ and $V_{\rm peak}$ are the flux density and velocity, respectively, of the maximum emission of each transition.  Flux density upper limits are at a 3$\sigma$ level. $\int S_\nu dV$ is the velocity-integrated flux density over the whole spectrum. The range of velocities over which emission is detected is given by $V_{\rm min}$ and $V_{\rm max}$. { Uncertainties of the flux density and the velocity-integrated flux density are at $1\sigma$ level. Velocities are given with respect to LSR.} 
}
\end{table*}

\subsubsection{IRAS 17239$-$2812}
\label{sec:17390}

Emission from the $^{28}$SiO transitions $v=1,2$, J=1-0 was detected by \citet{hal90} in single-dish observations toward the position listed in the IRAS point source catalog, using the Parkes Radio Telescope, with angular resolution of 1.6 arcmin, and { peak} flux densities 25 and 37 Jy for the $v=1$ and $v=2$ lines, respectively. { More recent interferometric observations with the Very Large Array (VLA) \citep{dic21} detected emission of the  $^{28}$SiO $v=0$, J=1-0 transition. Emission from this transition can have both thermal and maser origin, although \cite{dic21} conclude that, in this source, it is mostly of maser nature. No explicit coordinates for the emission is given by these authors, although their identification with a 2MASS source is consistent with ours. }

IRAS 17239$-$2812 can be associated with the infrared source  2MASS J17270541-2815299 and the optical object Gaia DR3 4059869891553026048, which provide a more accurate position. It has infrared colors compatible with those of AGB stars, and its OH spectrum shows the typical double-peak profile of this type of objects \citep{siv90}.
There are several reports of radio continuum emission, with catalog positions within the beam of the single-dish SiO observations, such as TXS 1723-282 \citep{dou96} at 365 MHz, MGPS J172702-281511 \citep{mur07} at 843 MHz, or NVSS	J172702-281525  \citep{con98} at 1.4 GHz. The formal position in the TXS catalog is only $\simeq 6''$  away from the optical/infrared counterpart of IRAS 17239$-$2812, a distance which is close to the positional uncertainty quoted for TXS 1723-282 ($4''$ in right ascension and $1.6''$ in declination). { However}, the positions of the sources reported in the MGPS and NVSS catalogs are significantly offset from the optical/infrared one, 
{ considering that their positional errors are}
better than $\simeq 1''-2''$ (e.g., NVSS J172702-281525 is $34''$ away from Gaia DR3 4059869891553026048 { but the position error reported for the radio source is $\simeq 0.09''\times 1.3''$}). 

\begin{figure*}
      \centering
      \sidecaption
            \includegraphics[width=0.7\hsize]{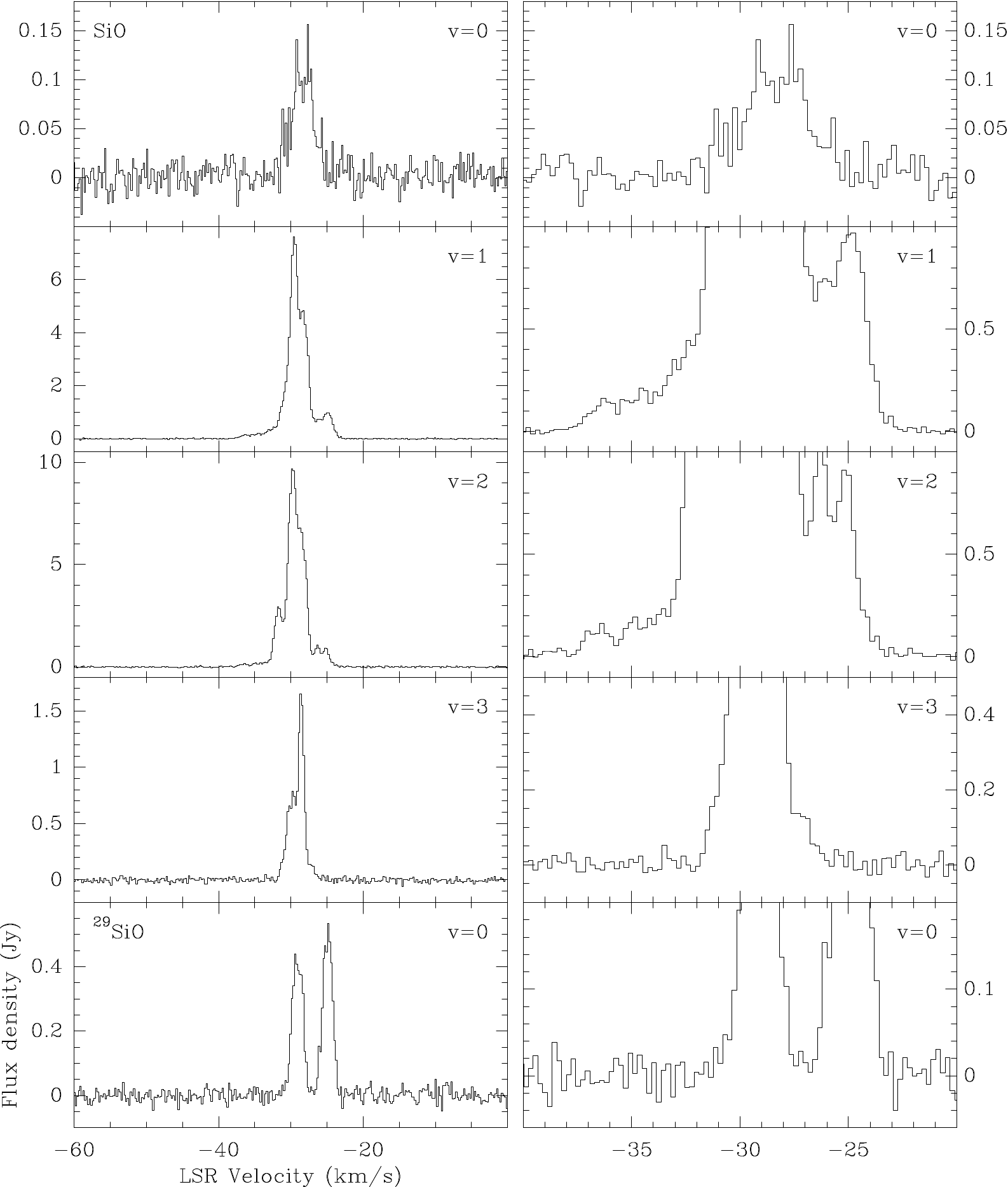}
      \caption{\label{fig:i17239_spec} SiO(J=1-0) spectra of IRAS 17239$-$2812. The spectra on the right column are the same as those in the left one { (no spectral smoothing applied)}, but zooming in on the x and/or y axes, to better show the weakest components. The spectra of the other objects presented in this paper are shown in Appendix \ref{app:figs}.}
\end{figure*}

\begin{figure}
      \centering
            \includegraphics[width=\hsize]{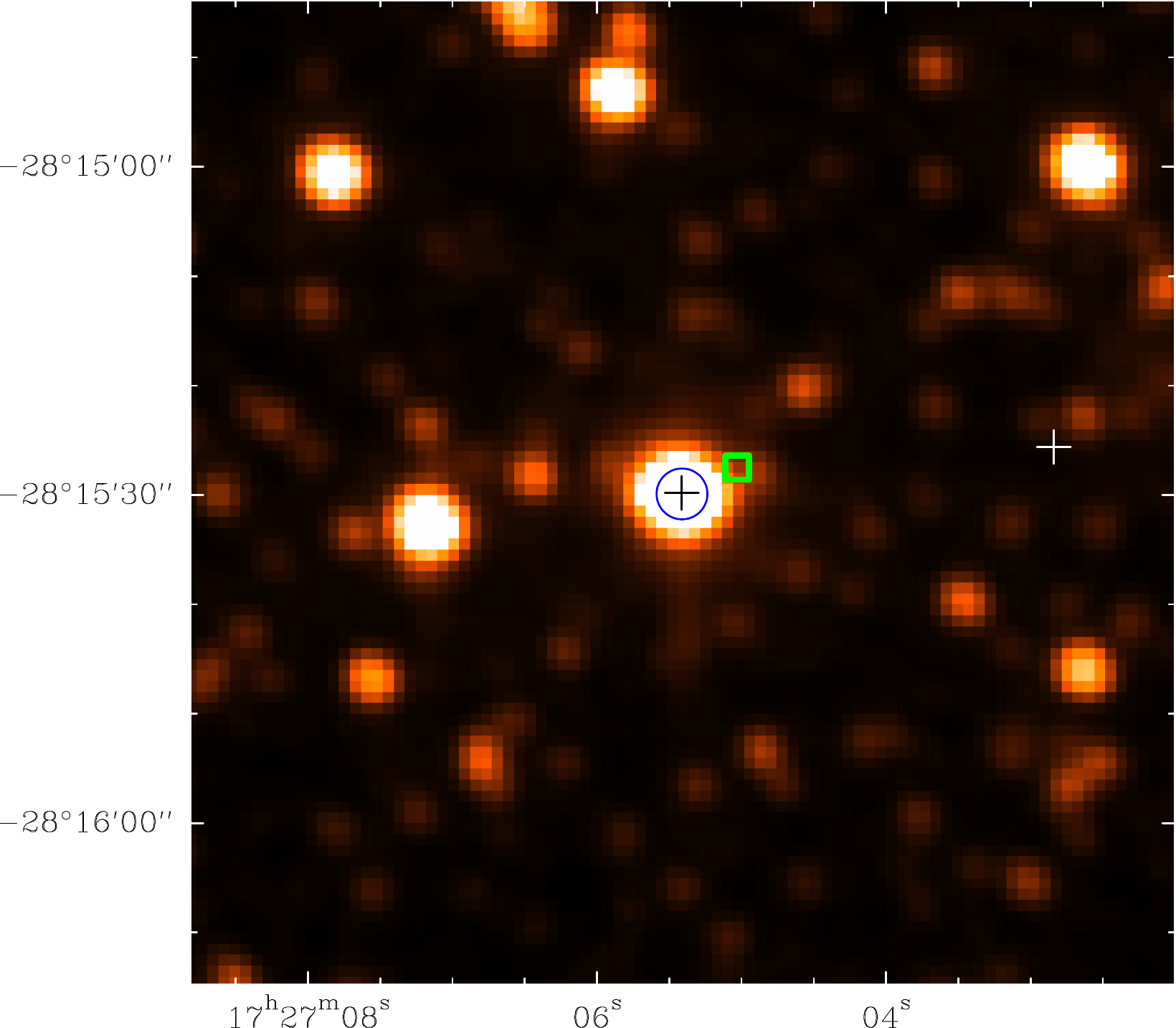}
      \caption{\label{fig:i17239_pos} 
      K-band infrared image from the 2MASS survey, centered at IRAS 17239$-$2812. The white cross marks the location of the radio source NVSS J172702-281525, the black cross is at the position of the { SiO maser emission} in our data, and the green square is at the catalog position of TXS 1723-282. The blue circle is centered at the optical source Gaia DR3 4059869891553026048. Coordinates are in the equatorial (equinox J2000) system.}
\end{figure}

We detected several transitions of SiO. Spectra are shown in Fig. \ref{fig:i17239_spec}. 
In Fig. \ref{fig:i17239_pos} we compare the location of the SiO { maser emission} with respect to  some of the the optical,
 infrared, and radio sources mentioned above. 
Our interferometric observations clearly show that the SiO emission is associated with IRAS 17239$-$2812, but we found no radio continuum emission, with $3\sigma$ upper limits of 90 and 130 $\mu$Jy beam$^{-1}$ at 43 and 45 GHz, respectively.
Moreover, the position of the SiO masers is not compatible with the radio continuum sources NVSS J172702-281525 nor MGPS J172702-281511. We also checked the radio continuum images at 2.1 GHz from the ATCA follow-up of the SPLASH survey \citep{qia16a}, processed by \citet{cal22}, and there is extended emission associated with NVSS J172702-281525, but no detection at the optical/infrared position of IRAS 17239$-$2812. It is also likely that the position listed for TXS 1723-282  has a larger error than the one reported in that catalog, and that the emission found in the TXS, MGPS, and NVSS catalogs actually corresponds to the same object. The possibility of such large positional errors in the TXS catalog is further discussed in \cite{dou96}. In any case, at this point we have no evidence that IRAS 17239$-$2812 may be associated with any radio continuum emission.

\subsubsection{IRAS 17390$-$3014}

$^{28}$SiO maser emission has been previously detected with the Nobeyama 45-m radio telescope by \cite{deg04} and \cite{fuj06}. { The reported values of peak flux density} for the J=1-0, $v=1$ and $v=2$ lines are similar in both works. For instance, \cite{deg04} obtain a peak flux density of $\simeq 1.4$ Jy for both transitions, while \cite{fuj06} get $\simeq 1.3$ Jy.

IRAS 17390$-$3014 is associated with the near-IR source 
2MASS J17421887-3015288 and the optical one
Gaia DR3 4056748416698878592. The OH maser spectrum shows a double-peaked profile typical of AGB stars \citep{sev97a,qia18}, with a separation between the peaks of $\simeq 23$ km s$^{-1}$, and a { central LSR velocity} of $\simeq -22$ km s$^{-1}$.
There is a radio continuum source at 1.4 GHz, NVSS J174218-301526 \citep{con98}, with a flux density of $10.2\pm 1.4$ mJy and a catalog position only $\simeq 6''$  away from that of the 2MASS and Gaia sources. Moreover, this radio source was reported to be extended, with  estimated deconvolved major and minor axes (full width at half maximum) of $66.2''\pm 12''$  and $<46.8''$, respectively, with position angle $6^\circ\pm 9^\circ$. Since the identification and parametrization of sources in the NVSS catalog \citep{con98} were carried out with an automated algorithm, applied on relatively large fields (size $4^\circ\times 4^\circ$), we downloaded the NVSS image to check the accuracy of the reported position and size estimates for the radio source.
We fitted a Gaussian with the image viewer of CASA, using a box that only contained NVSS J174218-301526, and obtained
deconvolved major and minor axes of $80''\pm 12''$ and $34''\pm 8''$  (position angle $-5^\circ\pm 6^\circ$), respectively, which are compatible with the values listed in the NVSS catalog. 
We will assume our size estimates in our calculations below.
The extended and elongated radio source contains the infrared/optical position of IRAS 17390$-$3014 so, in principle, it could be associated with this source, tracing, for instance a bipolar ionized nebula, as in many PNe. Moreover, both the infrared/optical and the radio positions are within the beam of the single-dish SiO observations, which makes IRAS 17390$-$3014 a tantalizing candidate for an SiO-maser-emitting PN.

Ours are the first reported interferometer observations of SiO masers in this region. The SiO spectra are shown in Fig. \ref{fig:i17390_spec}.
In Fig. \ref{fig:i17390_pos} we plot the relative positions of the { SiO maser emission} on an infrared image, where we have marked some of the sources mentioned in this section. No radio continuum emission has been detected in our data, with $3\sigma$ upper limits of 90 and 80 $\mu$Jy beam$^{-1}$ at 43 and 45 GHz, respectively. 
{In order to properly compare our limit with the detected radio source at 1.4 GHz (NVSS J174218-301526), we have to consider that the flux density is expected to increase, and the size of the emission to decrease at higher frequencies, in the case of nascent PNe. In quantitative terms,
if the radio source were tracing the free-free emission in a nascent PN, whose photoionized region is expected to have a radial dependency of particle density as $n\propto r^{-2}$, the flux density and linear size of the emission would vary with frequency as $S_\nu\propto \nu^{0.6}$ and $\theta_\nu\propto \nu^{-0.7}$,  respectively. These dependencies are valid both for a spherical ionized region \citep{pan75} and for a biconical, jet-like distribution with the same radial dependency of density \citep{rey86}. Therefore a source of 10.2 mJy at 1.4 GHz would emit $\simeq 80$ mJy at 43 GHz. Regarding its size,  assuming an ellipsoidal ionized region with a size dependence of $\theta_\nu\propto \nu^{-0.7}$ in all directions or a biconical distribution with that dependence along the major axis while maintaining the opening angle \citep{ang18},  its size would shrink from $80''\times 34''$ to $\simeq 7''\times 3''$. This translates to an expected mean intensity of $\simeq 0.5$ mJy beam$^{-1}$ at 43 GHz in our observations, which is well above the detection threshold of our continuum data (it should have been detected with a signal-to-noise ratio $\ga 15$).  Therefore, our non-detection in continuum at 43-45 GHz indicates that the continuum source cannot be tracing an ionized region with a radial dependence of $n\propto r^{-2}$, strongly suggesting that it is not associated with  a nascent PN. However, the continuum source could be tracing a more evolved PNe, with an optically thin  emission ($\alpha\simeq -0.1$) at least in part of the radio spectrum, and/or with a large size at 43-45 GHz, comparable with that at 1.4 GHz.}

To further consider the possibility that the radio emission is tracing a bipolar PN, physically associated with IRAS 17390$-$3014, we checked publicly available optical and infrared images at different wavelengths. In the images where IRAS 17390$-$3014 is detected, it shows up as a point source, with no hint of the extended nebulosity that could be expected if the radio continuum emission traces the ionized region of a PN. Therefore, we suggest that the radio continuum emission is not associated with IRAS 17390$-$3014, but it could be the superposition of several background sources within the $45''$ beam of the NVSS images. { It is, however, important to obtain sensitive radio continuum images with a resolution better than those $\simeq 45''$, to find out the sources that may be associated with that type of emission, and  to definitely ascertain whether or not some radio continuum emission could be actually related to IRAS 17390$-$3014. We checked the National Radio Astronomy Observatory Data Archive and the Australia Telescope Observational Archive, to search for publicly available VLA and ATCA observations, but we could not find any data with enough sensitivity for this goal.

\subsubsection{H 2-18 and IRAS 17403$-$2107}

\label{h2-18}

Our observations were centered at the radio position of the PN H 2-18 (Table \ref{tab:obs_atca}), reported by \cite{gat83}, with a positional accuracy of $\simeq 0.2''$. This radio source has a deconvolved size of $1.8''\pm 0.6''$ and a flux density of 11 mJy at 4.9 GHz. H 2-18 was first cataloged as a possible PN by \cite{har52}, and confirmed thereafter \citep[e.g.,][]{ack91,ack92}. It has a bipolar morphology in the optical, with a size of $\simeq 1.9''$  \citep{gue13}. { Its inclusion in our sample was based on a previous incorrect association of H 2-18 with} IRAS 17403-2107 in the SIMBAD database when we built our target list, while these two sources are actually 2.6 arcmin apart. This error was solved in SIMBAD at the time of submission of this paper. There are several single-dish SiO maser detections towards IRAS 17403-2107 \citep{izu95,deg07}, but no report of radio continuum emission. IRAS 17403-2107 is actually outside the primary beam of our interferometric observations. { This source has been classified as a Mira variable \citep{iwa22}.}

Even though H 2-18 does not really fulfill our criteria for being included in sample 2 (Sec. \ref{sec:sample2}) due to this incorrect { previous} association, we still think that reporting our results on this source is of interest in the context of searching for SiO emission in PNe. In our data,  we did not detect any SiO emission associated with the PN H 2-18. 

{ For further evidence of the distinction
between H2-18 and IRAS 17403-2107,  we point out}
that the counterparts of IRAS 17403-2107 at optical and near infrared wavelengths are Gaia DR3 4117388024866473088 and 2MASS J17431835-2109031, respectively, whereas H 2-18 is most likely associated with Gaia DR3 4117385040014126592 and 
2MASS  17432875-2109515, instead.

\subsubsection{IRAS 18052$-$2016 (OH 10.04$-$0.10)}

\label{sec:18052}

Emission of different transitions of 
$^{28}$SiO has been reported via single-dish observations, including the $v=1$, $J=2-1$ transition at 86 GHz \citep{dea07}, $v=1$ and $v=2$, $J=1-0$ at 43 GHz \citep{yoo14}. Maser emission of other molecular species, such as OH \citep[e.g.,][]{bau79a,sev97a} and H$_2$O \citep{eng86,yoo14}, has also been found.

IRAS 18052$-$2016 is associated with the near-IR source 
2MASS J18081638-2016115. There is no associated optical source in the Gaia DR3 catalog. 
Radio continuum emission has been reported in this source by \cite{gar88} and \cite{zoo90}, from simultaneous observations at 1441.5 and 1611.7 MHz with the VLA, with a flux density of 26 mJy in the combined dataset (central frequency = 1526.6 MHz). However, \cite{bec92} pointed out that the detected emission could actually be due to contamination from the OH maser, since it is only present in the band at 1611.7 MHz. { The bandpass of the VLA continuum data before 2010 was not sampled in frequency, which would have helped to identify the presence of spectral lines within the band}. Later observations by \cite{bai09} did not provide any detection of radio continuum emission, with reported $1\sigma$ noise values of 0.57 mJy beam$^{-1}$ at 4800 MHz (beam $=6.8''\times 2.6'')$ and  0.11 mJy beam$^{-1}$ at  8640 MHz (beam $=3.7''\times 1.4''$). In our data, we obtain a $3\sigma$ upper limit of 110 $\mu$Jy beam$^{-1}$ at both 43 and 45 GHz. We thus must conclude that there is no significant amount of ionized gas associated with this source. IRAS 18052$-$2016 could be a post-AGB star, based on the OH maser spectrum with four components \citep{qia20}, which departs from the usual double-peaked one in AGB objects, and on its infrared colors \citep{dea04}.

In Fig. \ref{fig:i18052_spec}, we show the SiO spectra we detected toward this source.
We note that the apparent absorption features around 55-60 km s$^{-1}$ in the $^{28}$SiO $v=0$ spectrum smoothed to 0.88 km s$^{-1}$ must be an artifact, since there is no background continuum emission against which line absorption can be produced.

\subsubsection{IRAS 19508+2659 (OH 63.9$-$0.2)}

\begin{figure}
      \centering
            \includegraphics[width=\hsize]{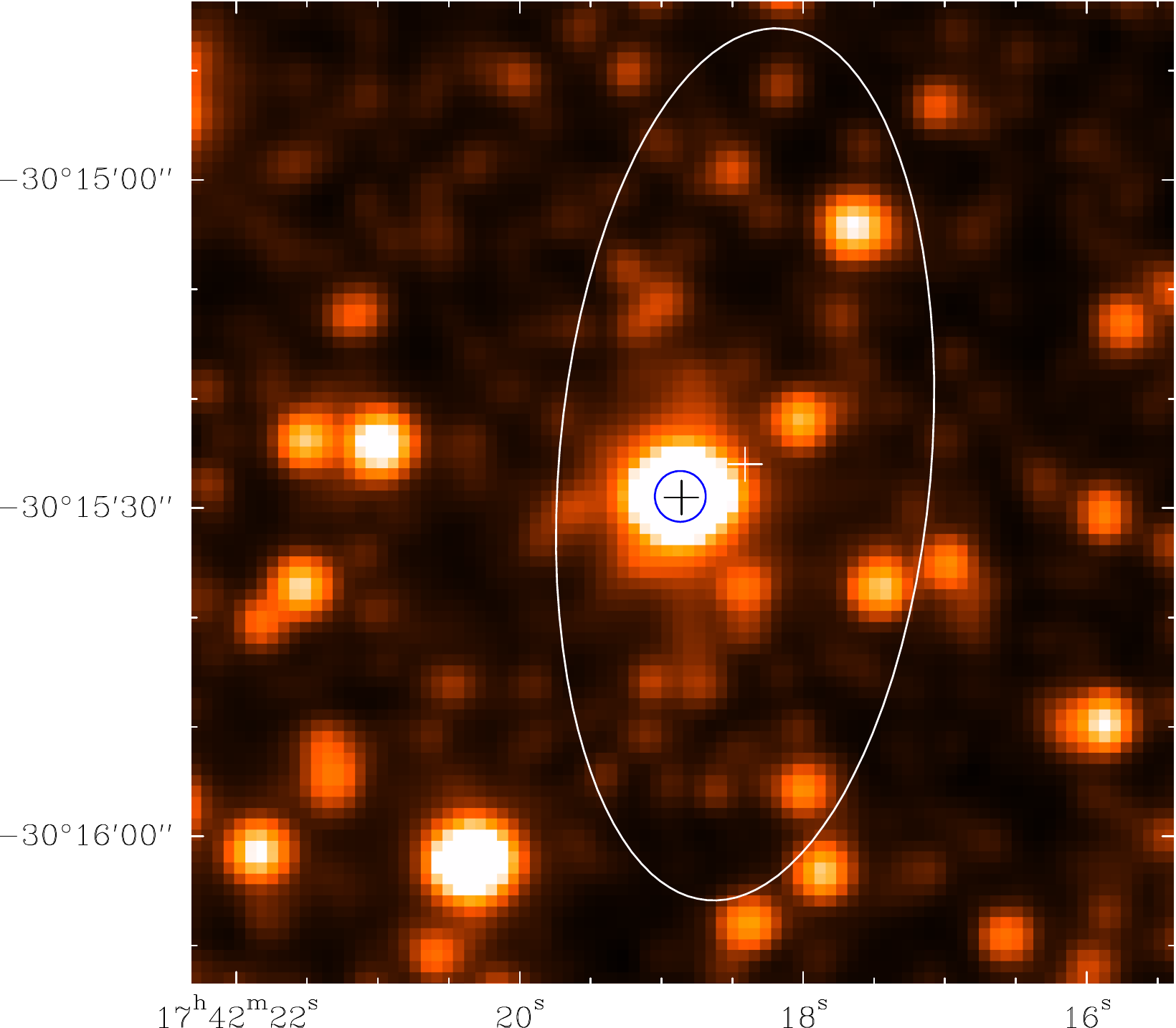}
      \caption{\label{fig:i17390_pos} 
      K-band infrared image from the 2MASS survey, centered at IRAS 17390$-$3014. 
      The white cross marks the center of the Gaussian fit to the radio source  NVSS J174218-301526, while the white ellipse represents the { full width at half maximum} size of this fit. The black cross is at the position of the { SiO maser emission} in our data. The blue circle is centered at the optical source Gaia DR3 4056748416698878592. Coordinates are in the equatorial (equinox J2000) system. }
\end{figure}

\label{sec:i19508}

Single-dish detections of $^{28}$SiO maser emission have been reported in this source by \cite{nak03a} and \cite{cho17}. 
The first single-dish report on OH maser was provided by \cite{bau79b}, and interferometric confirmation has been subsequently obtained \citep[THOR survey,][]{beu19}. The OH spectrum shows the typical double-peaked profile of { AGB} stars.
H$_2$O maser emission is also detected in this source  \citep{lew88,eng96}.

IRAS 19508+2659 is associated with sources 2MASS J19525786+2707447 and Gaia DR3 2027177220301507072. 
Radio continuum emission was reported by \cite{gar88} and \cite{zoo90} at 20 cm (mean frequency of 1526.6 MHz), with a flux density of $\simeq 13$ mJy. { However, note that this catalog has the problem (mentioned in section \ref{sec:18052} for IRAS 18052$-$2016) that one of the two bandpasses, centered at 1611.7 MHz and with a bandwidth of 3.125 MHz, includes the frequency of the OH maser line at 1612 MHz, so several sources in these catalogs were not really continuum sources, but were contaminated by OH maser emission. { \cite{bec92} compiled a list of sources in those data that showed emission at 1611.7 MHz but not at 1441.5 MHz and therefore, they were likely to be OH maser sources. IRAS 19508+2659 was not included in this list of OH maser sources.} Using the same data presented by \cite{gar88} and \cite{zoo90}, \cite{whi05} reported a value of continuum emission at 20 cm of $8.45	\pm 1.48$ mJy, at a position R.A.(J2000) $= 19^h52^m57.90^s$ (error = $0.7''$), Dec(J2000) $= +27^\circ07'44.7''$  (error = $0.8''$), and with a deconvolved major axis $\simeq 2.34''$. It is possible that these authors assumed that the continuum emission is real, and does not suffer from OH-maser contamination, since it is not in the \cite{bec92} catalog. However, we note that such an assumption  may not be correct in all cases, since the OH maser list of \cite{bec92} included only sources with a flux density $> 10$ mJy at 1611.7 MHz and undetected at 1441.5 MHz. So the exclusion from this list of the emission associated with IRAS 19508+2659 could just be because it fell below the emission threshold of 10 mJy in the 1611.7 MHz band.

To clarify whether or not IRAS 19508+2659 presents radio continuum emission, we downloaded the 20 cm data mentioned above (observed on 1983 December 30) from the VLA archive, and reprocessed them with the CASA package. In our images, we only detect emission at the position of IRAS 19508+2659 in the band centered at 1611.7 MHz, with a flux density of $10.1\pm 1.2$ mJy, and unresolved by the synthesized beam of the images ($6.5''\times 4.4''$). The $3\sigma$ upper limit to the emission at 1441.5 MHz is 3 mJy. Therefore, there is certainly OH maser contamination in these data. OH maser emission ar 1612 MHz has indeed been reported toward this source, in different single-dish observations \citep[e.g.,][]{win81,bau81}.

For a more complete check on the possible presence of radio continuum emission,
we have downloaded the NVSS images at 1.4 GHz, and found no radio source at this position. The NVSS image has a rms noise of $\simeq 0.8$ mJy beam$^{-1}$ (beam $=45''$), so it is sensitive enough to detect a source with the reported values of $8-13$ mJy. Moreover, no source at this position is listed in the catalog of radio continuum emission in the THOR survey \citep{wan18,wan20} at $1-2$ GHz, where they report $7\sigma$ upper limits of $\simeq 2-5$ mJy beam$^{-1}$ (beam $\simeq 10''-25''$). This limit should also have been enough to detect the previously reported flux densities. Considering that the free-free emission from ionized gas may be brighter at higher frequencies (unless it is already optically thin at 1.4 GHz, which is rare in planetary nebulae), 
we also downloaded the Very Large Array Sky Survey \citep[VLASS,][]{vlass}  images at 3 GHz, but we could not see any radio continuum associated with the source, with a $3\sigma$ upper limit of 3.6 mJy beam$^{-1}$ (beam = $3.1''\times 2.3''$). Moreover, recent VLA observations (Cala et al. in prep) did not provide a detection of the source at 22 GHz, with a $3\sigma$ upper limit of $\simeq 70$  $\mu$Jy beam$^{-1}$ (beam $\simeq 3.5''\times 2.6''$). In the data presented here, we did not detect any emission at 7 mm, with $3\sigma$ upper limits of 140 and 160 $\mu$Jy beam$^{-1}$ (beam $\simeq 1.13''\times 0.16''$ at 43 GHz), at 43 and 45 GHz, respectively. Therefore, we must conclude the absence of radio continuum emission from IRAS 19508+2659, with very stringent upper limits at different frequencies.

In Fig. \ref{fig:i19508_spec} we show the SiO spectra toward IRAS 19508+2659.
We only detect  $^{28}$SiO emission of the $v=1$ and $v=2$ transitions. 
 The position of the { SiO maser emission} and the emission at 1611.7 MHz (most likely arising from OH maser emission) are shown on an infrared image, in Fig. \ref{fig:i19508_pos}. Since there is no evidence for the presence of radio continuum emission, we then conclude that IRAS 19508+2659 is a regular AGB star.}

\begin{figure}
      \centering
            \includegraphics[width=\hsize]{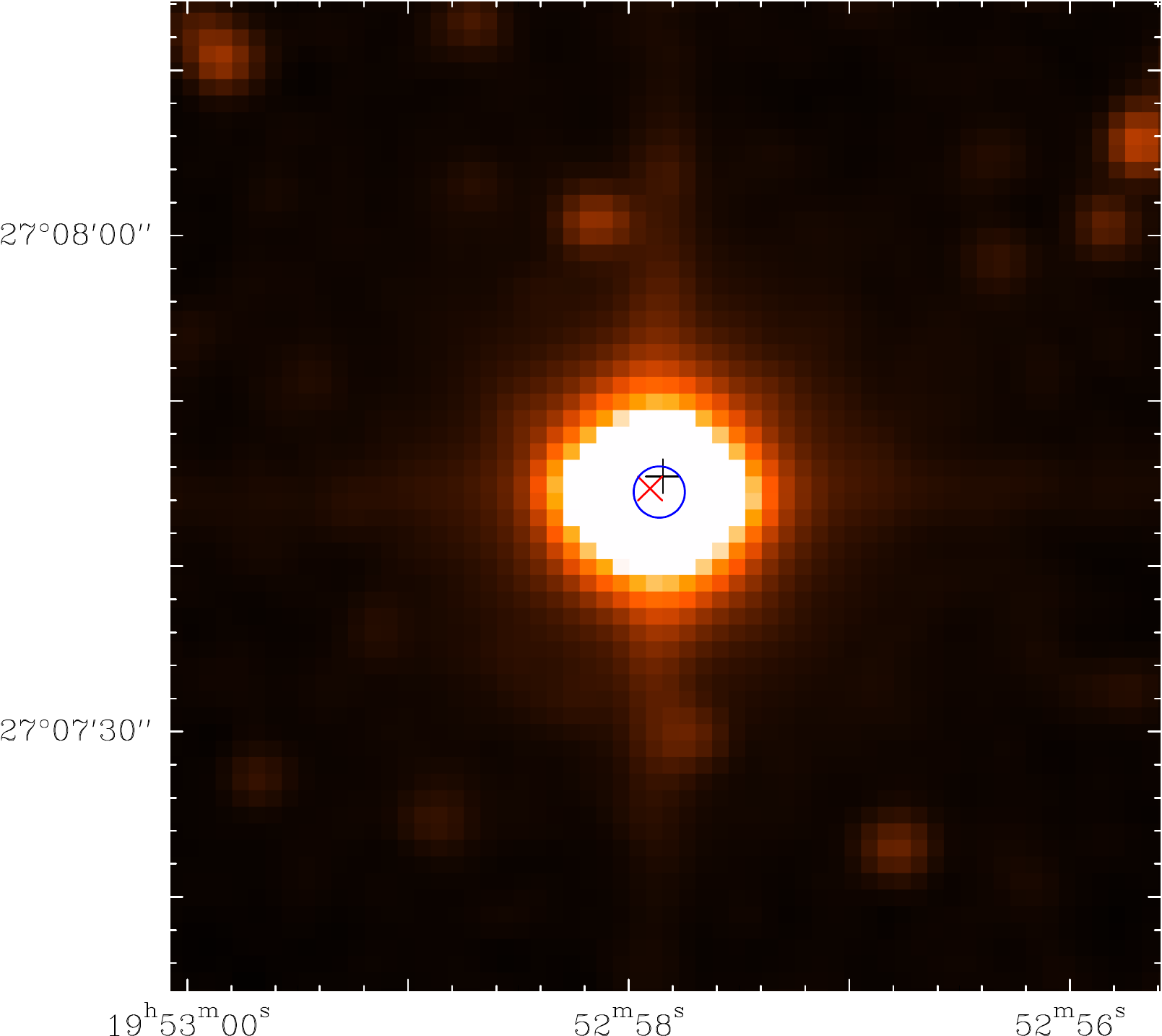}
      \caption{\label{fig:i19508_pos} 
      K-band infrared image from the 2MASS survey, centered at IRAS 19508+2659. The red cross marks the position of the radio continuum source reported by \cite{whi05}, { but that we argue it actually corresponds to OH maser emission (section \ref{sec:i19508})}.
      The black cross is at the position of the { SiO maser emission} in our data. The blue circle is centered at the optical source Gaia DR3 4056748416698878592. Coordinates are in the equatorial (J2000) system.  }
\end{figure}

\section{Discussion}
\label{sec:discussion}

The lack of confirmed association between SiO maser emission and PNe in this work has different implications, depending on the sample.
The negative results in sample 2 (section \ref{sec:sample2}) are due mainly to the fact that the reported radio continuum emission arises from a nearby, different source from that with SiO maser emission, and { therefore, there seems to be no SiO-maser-emitting PN within the beam of the previously reported single-dish maser observations}. However, 
our non-detection of SiO masers in nascent PNe (sample 1, section \ref{sec:sample1}) posses several interesting questions. 

One question is whether the abundance of SiO in gas phase in our targets is not enough to provide the necessary column density of molecules for detectable emission. Moreover, even if SiO is present in the circumstellar environment of PNe, it may be located in regions without the necessary physical conditions for population inversion of the molecular levels, which is a prerequisite to produce maser emission. Of course, it is also possible that SiO masers may be emitted in PNe different from the ones we sampled, specially considering that we targeted PNe harboring masers of other molecular species typical of O-rich stars (OH and H$_2$O).

 The SiO abundance in gas phase tends to drop in circumstellar envelopes as a function of distance from the star, because it is depleted onto dust grains \citep[e.g.,][]{buj86,ver19}. The SiO maser emission during the AGB phase is actually seen within $\la 5$ stellar radii, a distance shorter than the dust formation radius \citep{dia94}. However, SiO may be released again in the presence of shocks. This is the case, for instance, of the ``water fountain'' sources W43A (probably a late AGB star) and IRAS 16552-3050 (a post-AGB star), objects with clear high-velocity bipolar outflows \citep{sua08,taf20}, and which also show SiO maser emission \citep{nak03b,ama22}. In W43A, this emission seems to trace the base of the bipolar outflow, at distances $\ga 10$ au from the central star \citep{ima05},  larger than the radius of the SiO maser rings observed in typical AGB stars \citep[$<10$ au,][]{dia94}. { High-velocity outflows and shocks are also present in PNe \citep[e.g.,][]{had24,mir24}, and the detection of water maser emission in some of the objects of our sample 2 is itself a tracer of  shocks, since these are thought to be the main pumping mechanism of water masers. Therefore, one should reasonably expect that SiO molecules can be released from dust grains in some PNe.}
 
An additional question for PNe is whether the SiO molecules can survive photoionization close to the hot, central star. 
Thermal SiO emission at millimeter wavelengths has been reported in the bipolar PN M2-48, using single-dish observations \citep{edw14}.
{ Recent observations with the Atacama Large Millimeter Array of the nascent PN IRAS 15103-5754 (a source included in our sample 1) also show the presence of thermal SiO emission  \citep{gom23}}. This emission is found at distances $\la 400$ au from the central star. Thus, SiO molecules can indeed be found in gas phase around PNe, but they might not be at close enough distances to the star to produce maser emission. 

In the particular case of PNe showing OH and H$_2$O maser emission, which comprise our sample 1 (section \ref{sec:sample1}), we note that \cite{cal23} found some evidence that they have dual chemistry, with O-rich external envelopes and C-rich central stars. If the innermost regions of the central envelopes, where one could expect SiO masers to be pumped, are actually C-rich, then this emission will not be produced, due to a lack of SiO molecules. In this case, carbon-bearing molecules may be dominant in the gas phase near the central star.

In summary, the question of the possibility of SiO maser emission around PNe is still open. Our two carefully selected samples comprised what probably were the best candidates to date for such a detection, but none was obtained. { Our results clearly indicate that SiO masers, if present at all, are extremely rare.} A definite answer about the possibility of finding SiO masers in PNe, and firm constraints for non-detections would require large scale surveys and mappings of SiO maser emission in the Galactic plane and bulge. A promising example is the Bulge Asymmetries and Dynamical Evolution (BAaDE) survey \citep{lew21,sjo23}, which covers $\simeq 28,000$ sources, although it is still a targeted search, with a sample selected based on infrared colors. { 
The MALT-45 survey \citep{jor15} includes the SiO lines we studied in this paper, although its coverage of the galactic plane is limited, with published results only for galactic coordinates $330^\circ<l<335^\circ$, $|b| < 0.5^\circ$. }
A larger-scale blind SiO galactic mapping, as carried out for other maser species, like the Methanol Multibeam survey \citep{gre08}, the  Southern Parkes Large-Area Survey in Hydroxyl \citep[SPLASH,][]{daw22}, The HI/OH/Recombination line survey of the inner Milky Way \citep[THOR,][]{beu16}, or the H$_2$O southern Galactic Plane Survey \citep[HOPS,][]{wal11} may be necessary to avoid selection biases in the search for SiO-maser-emitting PNe.

\section{Conclusions}

{We have carried out a search for SiO maser emission in PNe. A total of 16 objects were observed in two source samples. We found no SiO maser emission associated with any confirmed or candidate PN. Our non-detections indicate that SiO masers in PN are extremely rare, if present at all. SiO molecules can be present in the gas phase in the circumstellar material of PN, but probably not at the locations needed for efficient pumping of SiO masers. Since there is some evidence that PN with OH and/or H$_2$O masers have O-rich outer envelopes, but C-rich central stars and inner envelopes, we speculate that C-bearing molecules can be dominant at the inner locations where the physical conditions for SiO maser production may have been met.}

\begin{acknowledgements}
The Australia Telescope Compact Array is part of the Australia Telescope National Facility (grid.421683.a) which is funded by the Australian Government for operation as a National Facility managed by CSIRO. We acknowledge the Gomeroi people as the traditional owners of the Observatory site. This work is financially supported by grants PID2020-114461GB-I00, PID2023-146295NB-I00, and CEX2021-001131-S, funded by MCIN/AEI /10.13039/501100011033, and by grant P20-00880, funded by the Economic Transformation, Industry, Knowledge and Universities Council of the Regional Government of Andalusia and the European Regional Development Fund from the European Union. RC also acknowledges support by the predoctoral grant PRE2018-085518, funded by MCIN/AEI/ 10.13039/501100011033 and by ESF Investing in your Future. 
We made use of the VLASS QLimage cutout server at URL cutouts.cirada.ca, operated by the Canadian Initiative for Radio Astronomy Data Analysis (CIRADA). CIRADA is funded by a grant from the Canada Foundation for Innovation 2017 Innovation Fund (Project 35999), as well as by the Provinces of Ontario, British Columbia, Alberta, Manitoba and Quebec, in collaboration with the National Research Council of Canada, the US National Radio Astronomy Observatory and Australia’s Commonwealth Scientific and Industrial Research Organisation. 
  
\end{acknowledgements}

\begin{appendix}

\section{Additional figures}
\label{app:figs}

In this appendix, we show the SiO(J=1-0) spectra of three of the sources detected in this paper: IRAS 17390$-$3014, IRAS 18052$-$2016, and IRAS 19508+2659. The spectra of IRAS 17239$-$2812 are presented in Fig. \ref{fig:i17239_spec}.

\begin{figure*}[]
      \centering
      \sidecaption
            \includegraphics[width=0.7\hsize]{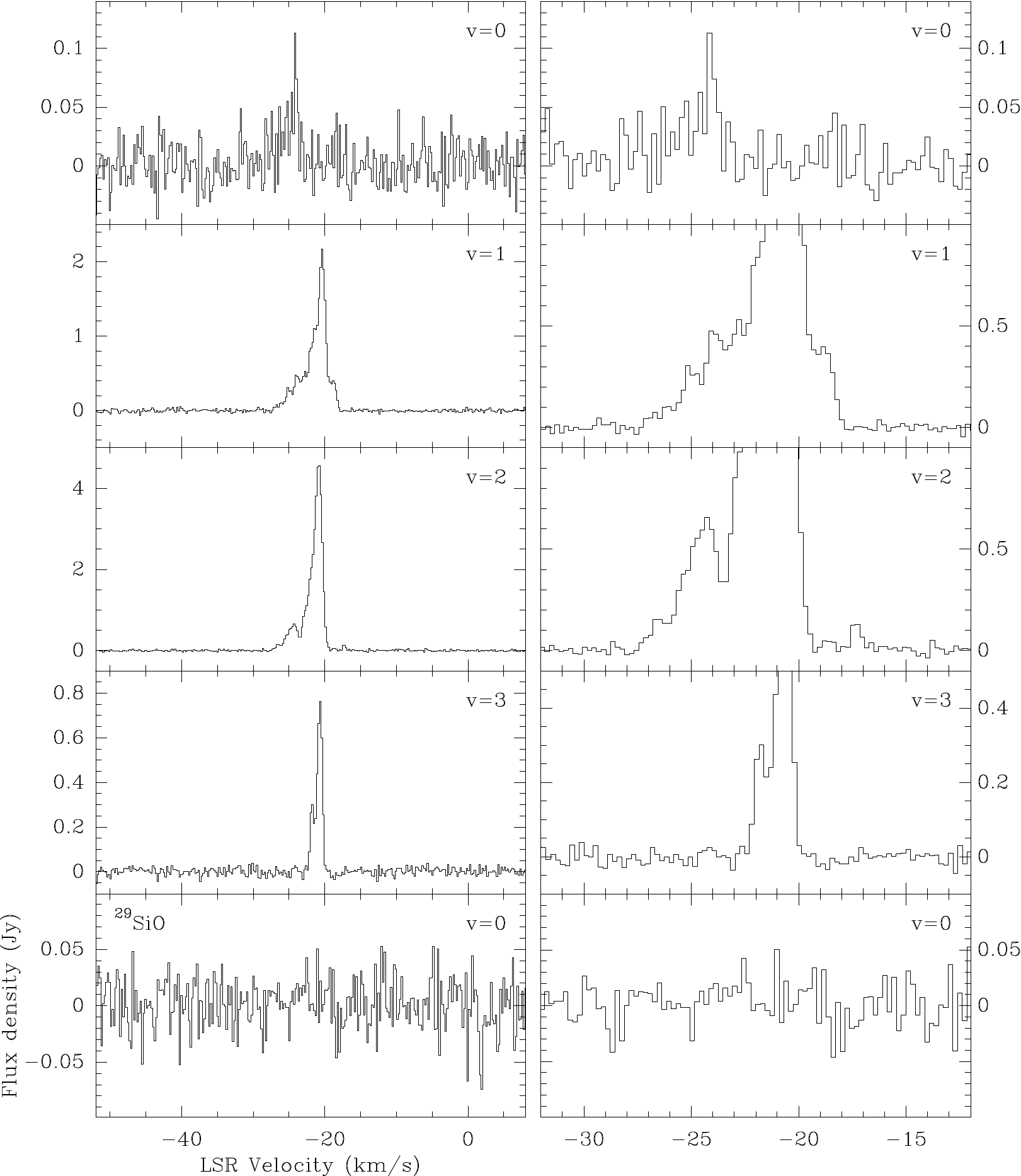}
      \caption{\label{fig:i17390_spec} Same as Fig. \ref{fig:i17239_spec}, but for IRAS 17390$-$3014.}
\end{figure*}

\begin{figure*}
      \centering
      \sidecaption
            \includegraphics[width=0.7\hsize]{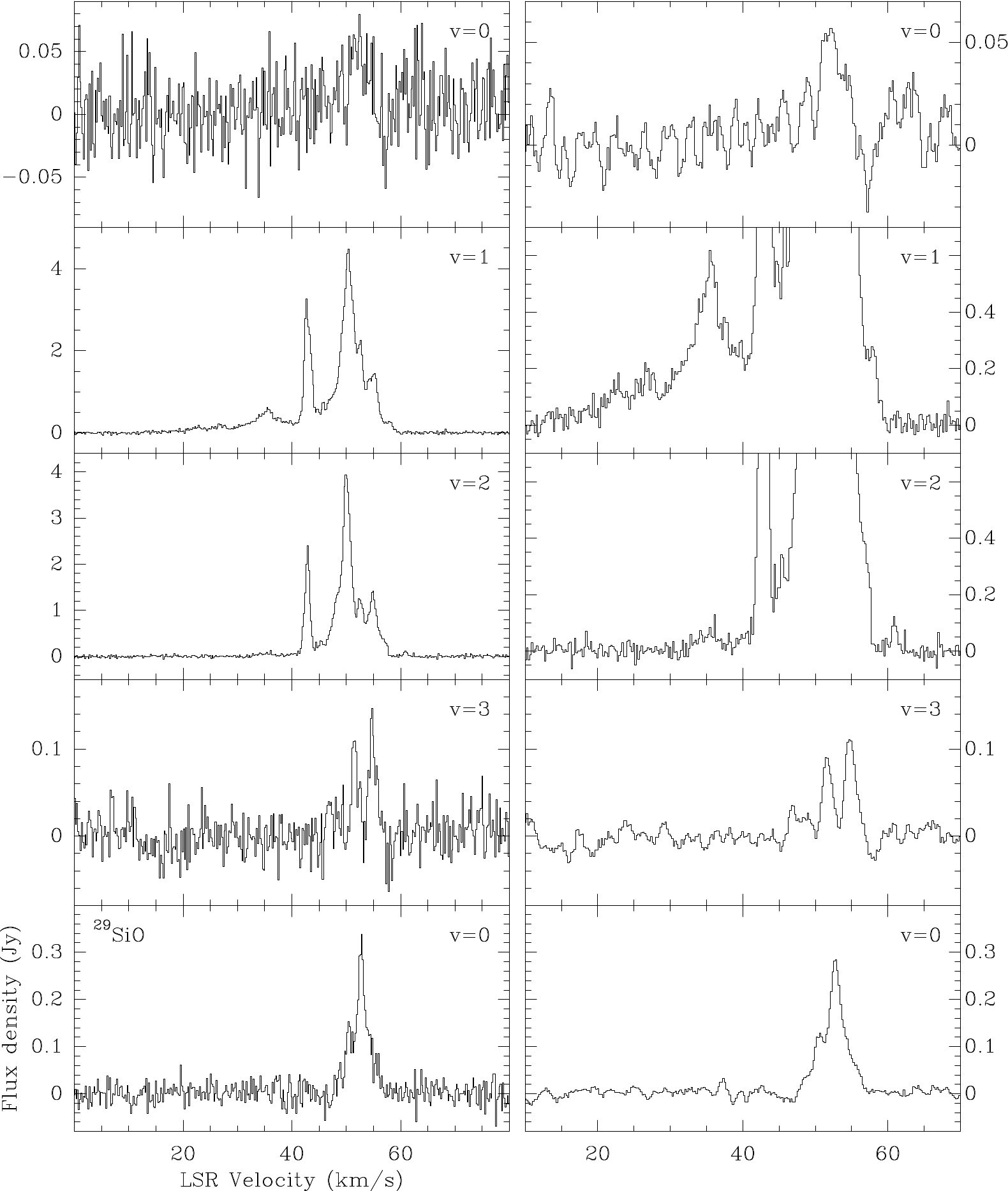}
      \caption{\label{fig:i18052_spec} SiO(J=1-0) spectra of IRAS 18052$-$2016. The spectra on the right column are the same as those in the left one, but zooming in on the x and/or y axes, to better show the weakest components, except in the case of the $v=0$ (both { isotopologues}) and $v=3$ lines, which have been { boxcar-smoothed} to a spectral resolution of 0.88 km s$^{-1}$, to increase the signal to noise ratio.}
\end{figure*}

\begin{figure*}
      \centering
      \sidecaption
            \includegraphics[width=0.7\hsize]{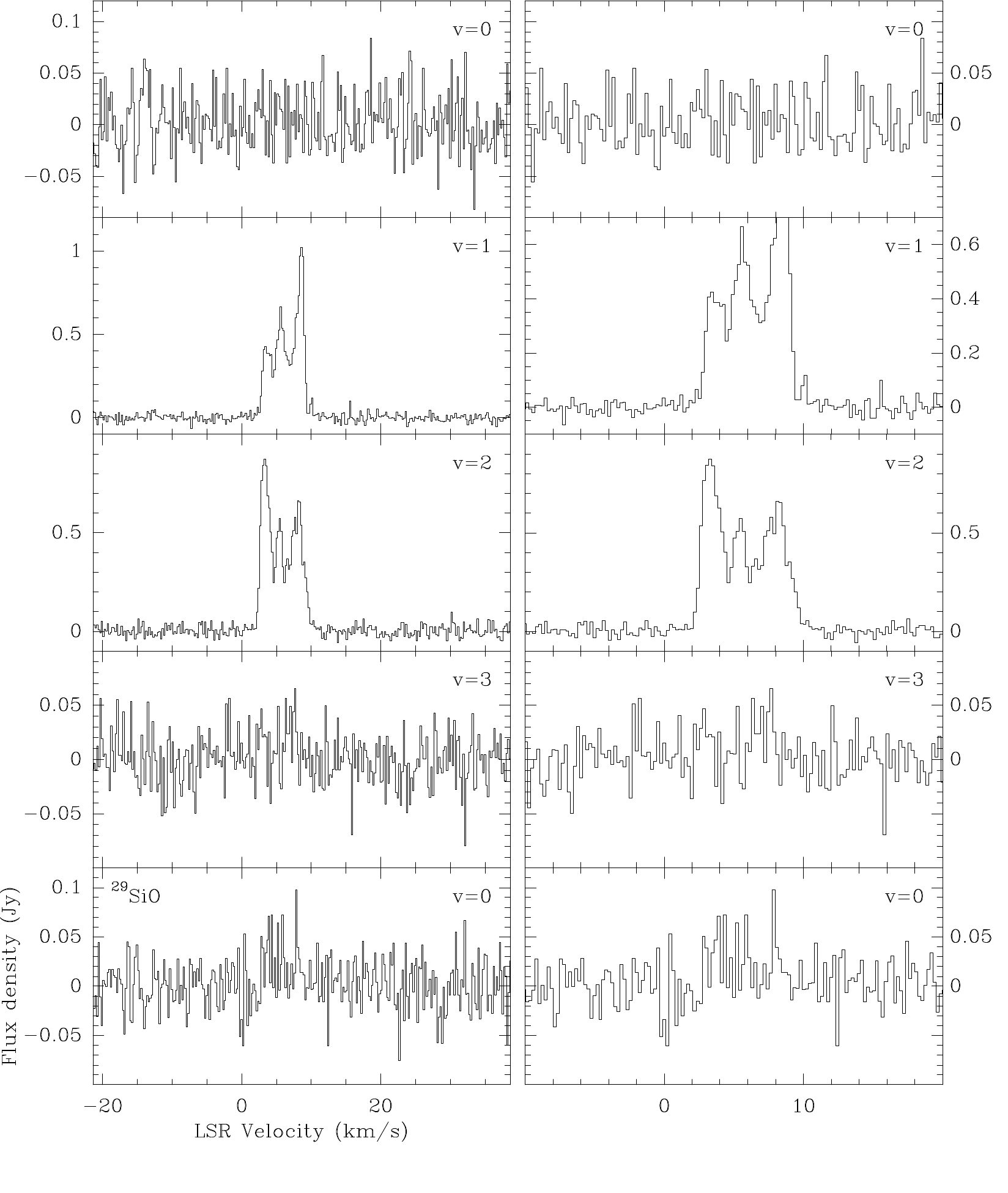}
      \caption{\label{fig:i19508_spec} Same as Fig. \ref{fig:i17239_spec}, but for IRAS 19508+2659. }
\end{figure*}

\end{appendix}


\begin{thebibliography}{}

  
\bibitem[Acker et al.(1992)]{ack92} Acker, A., Marcout, J., Ochsenbein, F., et al.\ 1992, The Strasbourg-ESO Catalogue of Galactic Planetary Nebulae. Parts I, II., European Southern Observatory, Garching (Germany), 1992
 
\bibitem[Acker et al.(1991)]{ack91} Acker, A., Raytchev, B., Koeppen, J., et al.\ 1991, \aaps, 89, 237

\bibitem[Amada et al.(2022)]{ama22} Amada, K., Imai, H., Hamae, Y., et al.\ 2022, \aj, 163, 85

\bibitem[Anglada et al.(2018)]{ang18} Anglada, G., Rodr{\'\i}guez, L.~F., \& Carrasco-Gonz{\'a}lez, C.\ 2018, \aapr, 26, 3

\bibitem[Bachiller et al.(1997)]{bac97} Bachiller, R., Forveille, T., Huggins, P.~J., et al.\ 1997, \aap, 324, 1123

\bibitem[Bains et al.(2009)]{bai09} Bains, I., Cohen, M., Chapman, J.~M., et al.\ 2009, \mnras, 397, 1386

\bibitem[Baud et al.(1979a)]{bau79a} Baud, B., Habing, H.~J., Matthews, H.~E., et al.\ 1979a, \aaps, 35, 179

\bibitem[Baud et al.(1979b)]{bau79b} Baud, B., Habing, H.~J., Matthews, H.~E., et al.\ 1979b, \aaps, 36, 193

\bibitem[Baud et al.(1981)]{bau81} Baud, B., Habing, H.~J., Matthews, H.~E., et al.\ 1981, \aap, 95, 156

\bibitem[Becker et al.(1992)]{bec92} Becker, R.~H., White, R.~L., \& Proctor, D.~D.\ 1992, \aj, 103, 544

\bibitem[Beuther et al.(2016)]{beu16} Beuther, H., Bihr, S., Rugel, M., et al.\ 2016, \aap, 595, A32

\bibitem[Beuther et al.(2019)]{beu19} Beuther, H., Walsh, A., Wang, Y., et al.\ 2019, \aap, 628, A90

\bibitem[Bl\"ocker(1995a)]{blo95a} Bl\"ocker, T.\ 1995a, \aap, 297, 727

\bibitem[Bl\"ocker(1995b)]{blo95b} Bl\"ocker, T.\ 1995b, \aap, 299, 755

\bibitem[Bowers \& Johnston(1994)]{bow94} Bowers, P.~F. \& Johnston, K.~J.\ 1994, \apjs, 92, 189

\bibitem[Bujarrabal et al.(1986)]{buj86} Bujarrabal, V., Planesas, P., Gomez-Gonzalez, J., et al.\ 1986, \aap, 162, 157

   \bibitem[Cala et al.(2022)]{cal22} Cala, R.~A., G{\'o}mez, J.~F., Miranda, L.~F., et al.\ 2022, \mnras, 516, 2235
   
     \bibitem[Cala et al.(2024)]{cal23} Cala, R.~A., G{\'o}mez, J.~F., Miranda, L.~F., et al.\ 2024, in Cosmic Masers: Proper Motion toward the Next-Generation Large Projects, IAU Symp 380, ed. T. Hirota, H. Imai, K. Menten, Y. Pihlstrom (Cambridge University Press, Cambridge), 343
     
     \bibitem[CASA Team et al.(2022)]{cas22} CASA Team, Bean, B., Bhatnagar, S., et al.\ 2022, \pasp, 134, 114501
     
     \bibitem[Caswell(1997)]{cas97} Caswell, J.~L.\ 1997, \mnras, 289, 203

  
   \bibitem[Cho et al.(2017)]{cho17} Cho, C.-Y., Cho, S.-H., Kim, S., et al.\ 2017, \apjs, 232, 13
   
   \bibitem[Clark(1980)]{cla80} Clark, B.~G.\ 1980, \aap, 89, 377

   \bibitem[Condon et al.(1998)]{con98} Condon, J.~J., Cotton, W.~D., Greisen, E.~W., et al.\ 1998, \aj, 115, 1693
   
   \bibitem[Dawson et al.(2022)]{daw22} Dawson, J.~R., Jones, P.~A., Purcell, C., et al.\ 2022, \mnras, 512, 3345
   
   \bibitem[Deacon et al.(2004)]{dea04} Deacon, R.~M., Chapman, J.~M., \& Green, A.~J.\ 2004, \apjs, 155, 595
   
   \bibitem[Deacon et al.(2007)]{dea07} Deacon, R.~M., Chapman, J.~M., Green, A.~J., et al.\ 2007, \apj, 658, 1096
   
   \bibitem[de Gregorio-Monsalvo et al.(2004)]{degr04} de Gregorio-Monsalvo, I., G{\'o}mez, Y., Anglada, G., et al.\ 2004, \apj, 601, 921

   \bibitem[Deguchi et al.(2004)]{deg04} Deguchi, S., Fujii, T., Glass, I.~S., et al.\ 2004, \pasj, 56, 765
   
   \bibitem[Deguchi et al.(2007)]{deg07} Deguchi, S., Fujii, T., Ita, Y., et al.\ 2007, \pasj, 59, 559
   
   \bibitem[Desmurs et al.(2000)]{des00} Desmurs, J.~F., Bujarrabal, V., Colomer, F., et al.\ 2000, \aap, 360, 189
   
   \bibitem[Diamond et al.(1994)]{dia94} Diamond, P.~J., Kemball, A.~J., Junor, W., et al.\ 1994, \apjl, 430, L61

	\bibitem[Dike et al.(2021)]{dic21} Dike, V., Morris, M.~R., Rich, R.~M., et al.\ 2021, \aj, 161, 111
	
	\bibitem[Douglas et al.(1996)]{dou96} Douglas, J.~N., Bash, F.~N., Bozyan, F.~A., et al.\ 1996, \aj, 111, 1945
	
\bibitem[Edwards \& Ziurys(2014)]{edw14} Edwards, J.~L. \& Ziurys, L.~M.\ 2014, \apjl, 794, L27

	\bibitem[Ellingsen et al.(2018)]{ell18} Ellingsen, S.~P., Voronkov, M.~A., Breen, S.~L., et al.\ 2018, \mnras, 480, 4851
\bibitem[Engels \& Jim{\'e}nez-Esteban(2007)]{eng07} Engels, D. \& Jim{\'e}nez-Esteban, F.\ 2007, \aap, 475, 941
	
	\bibitem[Engels \& Lewis(1996)]{eng96} Engels, D. \& Lewis, B.~M.\ 1996, \aaps, 116, 117
	
	\bibitem[Engels et al.(1986)]{eng86} Engels, D., Schmid-Burgk, J., \& Walmsley, C.~M.\ 1986, \aap, 167, 129
	
	\bibitem[Etoka \& Diamond(2010)]{eto10} Etoka, S. \& Diamond, P.~J.\ 2010, \mnras, 406, 2218
	
	\bibitem[Fomalont(1981)]{fom81} Fomalont, E.\ 1981, NRAO Newsletter, No. 3, P. 3, 1981, 3

\bibitem[Fujii et al.(2006)]{fuj06} Fujii, T., Deguchi, S., Ita, Y., et al.\ 2006, \pasj, 58, 529

\bibitem[Garc{\'\i}a-Hern{\'a}ndez et al.(2007)]{gar07} Garc{\'\i}a-Hern{\'a}ndez, D.~A., Perea-Calder{\'o}n, J.~V., Bobrowsky, M., et al.\ 2007, \apjl, 666, L33

\bibitem[Gaia Collaboration et al.(2016)]{gaia16} Gaia Collaboration, Prusti, T., de Bruijne, J.~H.~J., et al.\ 2016, \aap, 595, A1

\bibitem[Gaia Collaboration et al.(2023)]{gaia23} Gaia Collaboration, Vallenari, A., Brown, A.~G.~A., et al.\ 2023, \aap, 674, A1

\bibitem[Garwood et al.(1988)]{gar88} Garwood, R.~W., Perley, R.~A., Dickey, J.~M., et al.\ 1988, \aj, 96, 1655

\bibitem[Gathier et al.(1983)]{gat83} Gathier, R., Pottasch, S.~R., Goss, W.~M., et al.\ 1983, \aap, 128, 325

\bibitem[G{\'o}mez et al.(2024)]{gom23} G{\'o}mez, J.~F., Cala, R.~A.,  Miranda, L.~F., et al.\ 2024, in Cosmic Masers: Proper Motion toward the Next-Generation Large Projects, IAU Symp 380, ed. T. Hirota, H. Imai, K. Menten, Y. Pihlstrom (Cambridge University Press, Cambridge), 374

\bibitem[G{\'o}mez et al.(2005)]{gom05} G{\'o}mez, J.~F., de Gregorio-Monsalvo, I., Lovell, J.~E.~J., et al.\ 2005, \mnras, 364, 738

\bibitem[G{\'o}mez et al.(2008)]{gom08} G{\'o}mez, J.~F., Su{\'a}rez, O., G{\'o}mez, Y., et al.\ 2008, \aj, 135, 2074

\bibitem[G{\'o}mez et al.(2015)]{gom15} G{\'o}mez, J.~F., Su{\'a}rez, O., Bendjoya, P., et al.\ 2015, \apj, 799, 186

\bibitem[G{\'o}mez et al.(2016)]{gom16} G{\'o}mez, J.~F., Uscanga, L., Green, J.~A., et al.\ 2016, \mnras, 461, 3259

\bibitem[G{\'o}mez et al.(1990)]{gom90} G{\'o}mez, Y., Moran, J.~M., \& Rodr{\'\i}guez, L.~F.\ 1990, \rmxaa, 20, 55

\bibitem[Gordon et al.(2021)]{gor21} Gordon, Y.~A., Boyce, M.~M., O'Dea, C.~P., et al.\ 2021, \apjs, 255, 30

\bibitem[Green et al.(2008)]{gre08} Green, J.~A., Caswell, J.~L., Fuller, G.~A., et al.\ 2008, \mnras, 385, 948

\bibitem[Guerrero et al.(2013)]{gue13} Guerrero, M.~A., Toal{\'a}, J.~A., Medina, J.~J., et al.\ 2013, \aap, 557, A121

\bibitem[Hajduk et al.(2024)]{had24} Hajduk, M., van Hoof, P.~A.~M., Zijlstra, A.~A., et al.\ 2024, \aap, 688, L21

  \bibitem[Hall et al.(1990)]{hal90} Hall, P.~J., Wright, A.~E., Troup, E.~R., et al.\ 1990, \mnras, 247, 549
  
  \bibitem[Haro(1952)]{har52} Haro, G.\ 1952, Boletin de los Observatorios Tonantzintla y Tacubaya, 1, 1
  
  \bibitem[Henize(1967)]{hen67} Henize, K.~G.\ 1967, \apjs, 14, 125
  
  \bibitem[Imai et al.(2005)]{ima05} Imai, H., Nakashima, J.-. ichi ., Diamond, P.~J., et al.\ 2005, \apjl, 622, L125
  
  \bibitem[Iwanek et al.(2022)]{iwa22} Iwanek, P., Soszy{\'n}ski, I., Koz{\l}owski, S., et al.\ 2022, \apjs, 260, 46

\bibitem[Izumiura et al.(1995)]{izu95} Izumiura, H., Catchpole, R., Deguchi, S., et al.\ 1995, \apjs, 98, 271

\bibitem[Jacoby \& Van de Steene(2004)]{jast04} Jacoby, G.~H. \& Van de Steene, G.\ 2004, \aap, 419, 563

\bibitem[Jewell et al.(1991)]{jew91} Jewell, P.~R., Snyder, L.~E., Walmsley, C.~M., et al.\ 1991, \aap, 242, 211

\bibitem[Jordan et al.(2015)]{jor15} Jordan, C.~H., Walsh, A.~J., Lowe, V., et al.\ 2015, \mnras, 448, 2344

\bibitem[Kimeswenger(2001)]{kim01} Kimeswenger, S.\ 2001, \rmxaa, 37, 115

\bibitem[Lacy et al.(2020)]{vlass} Lacy, M., Baum, S.~A., Chandler, C.~J., et al.\ 2020, \pasp, 132, 035001

	\bibitem[Lewis \& Engels(1988)]{lew88} Lewis, B.~M. \& Engels, D.\ 1988, \nat, 332, 49
	

\bibitem[Lewis(2021)]{lew21} Lewis, M.~O.\ 2021, PhD thesis, University of New Mexico

\bibitem[Lovas \& Dragoset(2004)]{lov04} Lovas, F.~J., \& Dragoset, R.~A., in NIST Recommended Rest Frequencies for Observed Interstellar Molecular Microwave Transitions - 2002 Revision, (version 2.0.1). Available: http://physics.nist.gov/restfreq (2009, February 4). National Institute of Standards and Technology, Gaithersburg, MD. 

\bibitem[Menten et al.(2018)]{men08} Menten, K.~M., Wyrowski, F., Keller, D., et al.\ 2018, \aap, 613, A49

\bibitem[Miller Bertolami(2016)]{mil16} Miller Bertolami, M.~M.\ 2016, \aap, 588, A25

\bibitem[Miranda et al.(2001)]{mir01} Miranda, L.~F., G{\'o}mez, Y., Anglada, G., et al.\ 2001, \nat, 414, 284

\bibitem[Miranda et al.(1998)]{mir98} Miranda, L.~F., Torrelles, J.~M., Guerrero, M.~A., et al.\ 1998, \mnras, 298, 243

\bibitem[Miranda et al.(2024)]{mir24} Miranda, L.~F., V{\'a}zquez, R., Olgu{\'\i}n, L., et al.\ 2024, \aap, 687, A123

\bibitem[M\"uller et al.(2005)]{mul05} Müller, H.~S.~P., Schl\"oder, F., Stutzki, J. \&  Winnewisser, G. 2005,  J. Mol. Struct. 742, 215

  \bibitem[Murphy et al.(2007)]{mur07} Murphy, T., Mauch, T., Green, A., et al.\ 2007, \mnras, 382, 382
  
\bibitem[Nakashima \& Deguchi(2003a)]{nak03a} Nakashima, J.-I. \& Deguchi, S.\ 2003a, \pasj, 55, 203

\bibitem[Nakashima \& Deguchi(2003b)]{nak03b} Nakashima, J.-I. \& Deguchi, S.\ 2003b, \pasj, 55, 229

 \bibitem[Nyman et al.(1998)]{nym98} Nyman, L.-A., Hall, P.~J., \& Olofsson, H.\ 1998, \aaps, 127, 185
 
 \bibitem[Panagia \& Felli(1975)]{pan75} Panagia, N. \& Felli, M.\ 1975, \aap, 39, 1
 
 \bibitem[Payne et al.(1988)]{pay88} Payne, H.~E., Phillips, J.~A., \& Terzian, Y.\ 1988, \apj, 326, 368
 
 \bibitem[Pottasch et al.(1987)]{pot87} Pottasch, S.~R., Bignell, C., \& Zijlstra, A.\ 1987, \aap, 177, L49
  
  \bibitem[Qiao et al.(2020)]{qia20} Qiao, H.-H., Breen, S.~L., G{\'o}mez, J.~F., et al.\ 2020, \apjs, 247, 5

\bibitem[Qiao et al.(2018)]{qia18} Qiao, H.-H., Walsh, A.~J., Breen, S.~L., et al.\ 2018, \apjs, 239, 15

\bibitem[Qiao et al.(2016a)]{qia16a} Qiao, H.-H., Walsh, A.~J., Green, J.~A., et al.\ 2016a, \apjs, 227, 26

\bibitem[Qiao et al.(2016b)]{qia16b} Qiao, H.-H., Walsh, A.~J., G{\'o}mez, J.~F., et al.\ 2016b, \apj, 817, 37


\bibitem[Reid \& Menten(2007)]{rei07} Reid, M.~J. \& Menten, K.~M.\ 2007, \apj, 671, 2068

\bibitem[Reid et al.(1977)]{rei77} Reid, M.~J., Muhleman, D.~O., Moran, J.~M., et al.\ 1977, \apj, 214, 60

\bibitem[Reynolds(1986)]{rey86} Reynolds, S.~P.\ 1986, \apj, 304, 713.


\bibitem[Richards et al.(2012)]{ric12} Richards, A.~M.~S., Etoka, S., Gray, M.~D., et al.\ 2012, \aap, 546, A16

\bibitem[S{\'a}nchez Contreras et al.(2002)]{san02} S{\'a}nchez Contreras, C., Desmurs, J.~F., Bujarrabal, V., et al.\ 2002, \aap, 385, L1

\bibitem[Sault et al.(1995)]{sau95} Sault, R.~J., Teuben, P.~J., \& Wright, M.~C.~H.\ 1995, Astronomical Data Analysis Software and Systems IV, 77, 433

\bibitem[Seaquist \& Davis(1983)]{sea83} Seaquist, E.~R. \& Davis, L.~E.\ 1983, \apj, 274, 659

\bibitem[Sevenster et al.(1997a)]{sev97a} Sevenster, M.~N., Chapman, J.~M., Habing, H.~J., et al.\ 1997a, \aaps, 122, 79

\bibitem[Sevenster et al.(1997b)]{sev97b} Sevenster, M.~N., Chapman, J.~M., Habing, H.~J., et al.\ 1997b, \aaps, 124, 509

\bibitem[Sivagnanam et al.(1990)]{siv90} Sivagnanam, P., Le Squeren, A.~M., Minh, F.~T., et al.\ 1990, \aap, 233, 112

\bibitem[Sjouwerman(2023)]{sjo23} Sjouwerman, L.\ 2023, in Cosmic Masers: Proper Motion toward the Next-Generation Large Projects, IAU Symp 380, ed. T. Hirota, H. Imai, K. Menten, Y. Pihlstrom (Cambridge University Press, Cambridge), in press

\bibitem[Spitzer Science(2009)]{spitzer} Spitzer Science, C.\ 2009, VizieR Online Data Catalog, 2293. II/293

\bibitem[Skrutskie et al.(2006)]{2mass} Skrutskie, M.~F., Cutri, R.~M., Stiening, R., et al.\ 2006, \aj, 131, 1163. doi:10.1086/498708

\bibitem[Stevens(2022)]{ste22} Stevens J., Wark R., Edwards P., et al.\ 2015,
ATCA Users Guide, Revision 1.10, http://www.narrabri.atnf.csiro.au/observing/users\_
guide/html/atug.html

\bibitem[Su{\'a}rez et al.(2006)]{sua06} Su{\'a}rez, O., Garc{\'\i}a-Lario, P., Manchado, A., et al.\ 2006, \aap, 458, 173

\bibitem[Su{\'a}rez et al.(2015)]{sua15} Su{\'a}rez, O., G{\'o}mez, J.~F., Bendjoya, P., et al.\ 2015, \apj, 806, 105

\bibitem[Su{\'a}rez et al.(2008)]{sua08} Su{\'a}rez, O., G{\'o}mez, J.~F., \& Miranda, L.~F.\ 2008, \apj, 689, 430

\bibitem[Su{\'a}rez et al.(2007)]{sua07} Su{\'a}rez, O., G{\'o}mez, J.~F., \& Morata, O.\ 2007, \aap, 467, 1085

\bibitem[Tafoya et al.(2009)]{taf09} Tafoya, D., G{\'o}mez, Y., Patel, N.~A., et al.\ 2009, \apj, 691, 611

\bibitem[Tafoya et al.(2020)]{taf20} Tafoya, D., Imai, H., G{\'o}mez, J.~F., et al.\ 2020, \apjl, 890, L14

\bibitem[te Lintel Hekkert et al.(1991)]{tel91} te Lintel Hekkert, P., Caswell, J.~L., Habing, H.~J., et al.\ 1991, \aaps, 90, 327

\bibitem[Turner(1979)]{tur79} Turner, B.~E.\ 1979, \aaps, 37, 1

\bibitem[Uscanga et al.(2012)]{usc12} Uscanga, L., G{\'o}mez, J.~F., Su{\'a}rez, O., et al.\ 2012, \aap, 547, A40

\bibitem[Uscanga et al.(2014)]{usc14} Uscanga, L., G{\'o}mez, J.~F., Miranda, L.~F., et al.\ 2014, \mnras, 444, 217

\bibitem[Vassiliadis \& Wood(1993)]{vas93} Vassiliadis, E. \& Wood, P.~R.\ 1993, \apj, 413, 641

\bibitem[Vassiliadis \& Wood(1994)]{vas94} Vassiliadis, E. \& Wood, P.~R.\ 1994, \apjs, 92, 125

\bibitem[Verbena et al.(2019)]{ver19} Verbena, J.~L., Bujarrabal, V., Alcolea, J., et al.\ 2019, \aap, 624, A107

\bibitem[Vyssotsky et al.(1945)]{vys45} Vyssotsky, A.~N., Miller, W.~J., \& Walther, M.~E.\ 1945, \pasp, 57, 314

\bibitem[Walsh et al.(2011)]{wal11} Walsh, A.~J., Breen, S.~L., Britton, T., et al.\ 2011, \mnras, 416, 1764

\bibitem[Wang et al.(2018)]{wan18} Wang, Y., Bihr, S., Rugel, M., et al.\ 2018, \aap, 619, A124

\bibitem[Wang et al.(2020)]{wan20} Wang, Y., Bihr, S., Rugel, M., et al.\ 2020, \aap, 641, C1

\bibitem[White et al.(2005)]{whi05} White, R.~L., Becker, R.~H., \& Helfand, D.~J.\ 2005, \aj, 130, 586

\bibitem[Winnberg et al.(1981)]{win81} Winnberg, A., Terzides, C., \& Matthews, H.~E.\ 1981, \aj, 86, 410

\bibitem[Yoon et al.(2014)]{yoo14} Yoon, D.-H., Cho, S.-H., Kim, J., et al.\ 2014, \apjs, 211, 15

\bibitem[Zijlstra et al.(1989)]{zij89} Zijlstra, A.~A., te Lintel Hekkert, P., Pottasch, S.~R., et al.\ 1989, \aap, 217, 157

\bibitem[Zoonematkermani et al.(1990)]{zoo90} Zoonematkermani, S., Helfand, D.~J., Becker, R.~H., et al.\ 1990, \apjs, 74, 181

\end{thebibliography}
\end{document}